\documentclass{amsart}

\usepackage{amssymb}
\usepackage{amsmath}
\usepackage{bbm}
\usepackage{graphicx}
\def\img{\includegraphics}

\newtheorem{theorem}{Theorem}[section]
\newtheorem{lemma}[theorem]{Lemma}
\newtheorem{proposition}[theorem]{Proposition}
\newtheorem{corollary}[theorem]{Corollary}

\theoremstyle{definition}

\theoremstyle{remark}
\newtheorem{remark}[theorem]{Remark}

\numberwithin{equation}{section}

\newcommand{\assign}{:=}

\newcommand{\tmmathbf}[1]{\ensuremath{\boldsymbol{#1}}}
\newcommand{\tmop}[1]{\ensuremath{\operatorname{#1}}}
\newcommand{\tmstrong}[1]{\textbf{#1}}
\newenvironment{itemizedot}{\begin{itemize} }{\end{itemize}}

\def\Res{{\tmop{Res}}}
\def\SRes{\mathrm{SRes}^{(1)}}
\def\Disc{{\mathrm{Disc}}}
\def\mm{{\mathfrak{m}}}
\def\Dm{{\mathfrak{D}}}
\def\Fm{{\mathfrak{F}}}
\def\Rm{{\mathfrak{R}}}
\def\Tm{{\mathfrak{T}}}
\def\Um{{\mathfrak{U}}}

\def\NN{{\mathbb{N}}}
\def\RR{{\mathbb{R}}}
\def\SS{{\mathbb{S}}}
\def\ZZ{{\mathbb{Z}}}
\def\QQ{{\mathbb{Q}}}
\def\UU{{\mathbb{U}}}
\def\Rc{{\mathcal{R}}}
\def\Dc{{\mathcal{D}}}

\begin{document}

\title{Explicit factors of some iterated resultants and discriminants}
\thanks{This work was first presented at the conference in honor of Jean-Pierre Jouanolou, held at Luminy, Marseille, 15-19$^{\rm th}$ May 2006}

\author[L.~Bus\'e]{Laurent Bus\'e}
\address{GALAAD, INRIA, B.P.~93, 06902 Sophia Antipolis, France}
\email{Laurent.Buse@inria.fr}

\author[B.~Mourrain]{Bernard Mourrain}
\address{GALAAD, INRIA, B.P.~93, 06902 Sophia Antipolis, France}
\email{mourrain@sophia.inria.fr}


\date{\today}

\dedicatory{dedicated to Professor Jean-Pierre Jouanolou}

\begin{abstract}
In this paper, the result of applying iterative univariate
resultant constructions to multivariate polynomials is analyzed. We consider the input
polynomials as generic polynomials of a given degree and exhibit explicit
decompositions into irreducible factors of several constructions involving
two times iterated univariate resultants and discriminants over the integer
universal ring of coefficients of the entry polynomials. Cases
involving from two to four generic polynomials and resultants or
discriminants in one of their variables are treated. The decompositions into irreducible 
factors we get are obtained by exploiting fundamental properties of the univariate
resultants and discriminants and induction on the degree of the polynomials. As a consequence,  
each irreducible factor can be separately and explicitly 
computed in terms of a certain multivariate resultant. With this approach, we also obtain as 
direct corollaries some results conjectured by Collins \cite{Collins} and
McCallum \cite{Callum,Calpreprint} which correspond to the case of polynomials
whose coefficients are themselves generic polynomials in other 
variables. Finally, a geometric interpretation of the algebraic factorization
of the iterated discriminant of a single polynomial is detailled. 
\end{abstract}

\maketitle

\section{Introduction}

Resultants provide an essential tool in constructive algebra and
in equation solving, for projecting the solution of a polynomial system into
a space of smaller dimension.
In the univariate case, a well-known construction due to 
J.J. Sylvester (1840) consists in eliminating the monomials
$1,X,\ldots,X^{m+n-1}$ in the multiples
$$
(X^{i}P(X))_{0\leq i\leq n-1}, (X^{j}Q(X))_{0 \leq j\leq m-1},
$$
of two given polynomials $P,Q$ of degree $m$ and $n$, and in taking the 
determinant of the corresponding $(m+n)\times(m+n)$ matrix.
Though the first resultant construction appeared probably in the work of
E. B\'ezout \cite{Bez79} (see also Euler's work), and although 
contemporary to related works (Jacobi 1835, Richelot 1840, Cauchy
1840, \dots), this method remains well-known as Sylvester's resultant.
It is nowadays a fundamental tool used in effective algebra to
eliminate a variable between two polynomials.

It is a natural belief that equipped with such a tool which eliminates one
variable at a time, one can iteratively  eliminate several variables. 
This approach was actually exploited, for instance in \cite{VdW}, to deduce
theoretical results (such as the existence of eliminant polynomials) in
several variables.  However, if we are interested in structural results as
well as practical computations
or complexity issues, this approach is far from being optimal. 
This explains the study and development of different types of multivariate
resultants, including projective \cite{Mac02, Jou91}, anisotropic
\cite{Jou91,Jou96}, toric \cite{GKZ,emiris95,Carlos}, residual 
\cite{Jou96,BusPhD,BEM01}, determinantal \cite{Bus04} resultants. 

Nevertheless, in some algorithms, such as in Cylindrical Algebraic
Decomposition (CAD), an induction is applied on the dimension of the problems
and iterated univariate resultants and subresultants are used at many steps
of the algorithm \cite{Collins,bpr-arag-03}. In its original paper on quantifier
elimination for real closed fields by cylindrical algebraic decomposition
\cite{Collins}, Collins used these iterated resultants in a geometric context
and observed certain intriguing factorizations that ``suggest some
theorems'' \cite[p.178]{Collins}. The same year, 1975, Van der Waerden
responded to these observations in a handwritten letter where he gave some
intuitive hints for some of the phenomena noticed by Collins. 
More recently, McCallum \cite{Callum,Calpreprint} proved rigorously that certain iterated
resultants have some irreducible factors, but he only showed the
existence and did not give a way to compute them independently. As pointed out by 
Jean-Pierre Jouanolou, in 1868 and 1869  Olaus Henrici published
two outstanding papers \cite{He1868,He1869} 
addressing the decomposition of the discriminant of a discriminant. In particular,
 he gave the expected factorization of such a repeated discriminant. However, 
he did not prove the irreducibility of these factors which is a more difficult task. 
One of our goals in this paper is to give the decomposition into irreducible 
factors of these iterated resultant computations.

From a geometric point of view, we are interested in the solutions of
equations depending on some parameters and by analyzing what happens
``above'', when we move these parameters. The number of solutions might
change if we cross the set of points where a vertical line is tangent to the
solution set (that is to the polar variety in the vertical direction). 
This polar variety projected in one dimension less, might have singularities where
the number of solutions changes effectively or which are only due to the
superposition of distinct points of this polar variety. 
These critical points of the polar variety of an algebraic surface
$f(x,y,z)=0$ in $\RR^{3}$ are used effectively in algorithms for computing
the topology of the surface (see e.g.~\cite{mt:imras-05}).  The projection of
the polar curve (say on the $(x,y)$-plane) is obtained by a discriminant
computation and the critical points of this projected curve are again
computed by a discriminant. These values are then used to
analyze where the topology of a plane section is changing in order to deduce
the topology of the whole surface $f(x,y,z)=0$. Similar projection tools are
also implicitly used in higher dimension for the triangulation of
hypersurfaces (see e.g. \cite{Hardt1}), which leads in the algebraic context
to univariate resultant computation (see e.g. \cite{Co-isg-02, bpr-arag-03}).

Another of our objectives is to show how these critical points corresponding
generically to folds, double folds or pleats of the surface can be related to
explicit factors in iterated resultant constructions.  The main results of
this paper are complete explicit factorizations of two times iterated
univariate resultants and discriminants of generic polynomials of given
degree. We actually give the decomposition of these iterated resultants over
the integer universal ring of coefficients of polynomials of given
degree. Such a formulation has the advantage to allow the pre-computation of
a given factor. It has direct applications to the topological computation of
algebraic surfaces, which was our starting point.

Our approach is based on the study of these iterated resultants in 
generic situations. Most of the interesting formulas are obtained by
suitable specialization of this case. These specializations are performed
using the formalism of the multivariate resultant as it has been originaly introduced and deeply developed
by Jouanolou to whom this paper is dedicated in recognation of its outstanding contributions to resultant theory (e.g.~\cite{Jou91,Jou96,Jou97}). It should be noticed that this approach can be
pushed further to study more particular situations (corresponding to other
type of multivariate resultants) that we did not consider in
this paper, but which could be interesting for specific applications.

The paper is structured as follows. In the next section, we recall the 
definitions and main properties of resultants and discriminants that we will
use. In section 3, we consider the computation of iterated resultants of $4$
and then $3$ polynomials. In section 4, we analyze the iterated computation of resultants of 
discriminants, first the resultant of the discriminants of two distinct
polynomials $P_{1}$, $P_{2}$ and next the resultant of the discriminant of $P_{1}$ and the
resultant of $P_{1}, P_{2}$. We here extend the previous work \cite{Callum} and prove 
some properties conjectured by Collins and McCallum.
In section 5, we study the discriminant of a resultant, simplifying the proof
and also extending some results of \cite{Callum,Calpreprint}.
These developments are used in cascade to provide, in section 6, the complete
factorization into irreducible  components of a discriminant of a
discriminant for a generic polynomial, as conjectured in \cite{Calpreprint}.
These new results have direct corollaries for polynomials whose coefficients
are themselves generic polynomials in other parameter variables, which we provide.

\section{Background material and notation}

In this section we give the notation and  quickly present the tools, as
resultants and discriminants, that we will use all along this paper. 
 
\subsection{Resultants}

\subsubsection{The univariate case}
Let $\SS$ be a commutative ring (with unity) and consider the two 
polynomials in $\SS[X]$
$$
f(X)  :=  a_mX^m+a_{m-1}X^{m-1}+\cdots+a_0,\  
g(X)  :=  b_nX^n+b_{n-1}X^{n-1}+\cdots+b_0,
$$
where $a_{i}, b_{j} \in \SS$, and $m$ and $n$ are both positive 
integers. Their resultant (in degrees $m,n$)\footnote{Notice that the dependence on
the degrees $m,n$ can be avoided if one considers homogeneous
polynomials.} that we will denote $\Res_X(f,g)$, is defined as the
determinant of the well-known Sylvester matrix 
$$\left(\begin{array}{ccccccc}
  a_m & 0 & \cdots & 0 & b_n & 0 & 0 \\
  a_{m-1} & a_m &      & \vdots & b_{n-1} & \ddots &    0   \\
  \vdots &  & \ddots &  0  & \vdots &  & b_n      \\
  a_0 &  &  & a_m & b_{1} &  & b_{n-1}  \\ 
  0 & a_m  &  & a_{m-1} & b_{0} &  &   \vdots  \\
 \vdots&   & \ddots  & \vdots & 0 & \ddots & b_{1}    \\
   0 &  \cdots & 0  & a_0 & 0 & 0  &   b_0  \\
\end{array}\right).$$
\begin{remark}\label{ResNot} We emphasize that the notation $\Res_X(f,g)$ denotes the resultant of $f$ and
$g$ with respect to the variable $X$ as polynomials of their \emph{expected}
degree, which is here $m$ and $n$ respectively. It is important to keep this
in mind since the closed formulas we will prove
in this paper are using this convention; see Remark \ref{Resmn} for an
illustration.
\end{remark}
 
This ``eliminant polynomial'' has a long history and many known
properties. One can learn about it in many places in the literature, for
instance \cite{CLO,Lang}; see also \cite[chapter 12]{GKZ} and \cite{ApJo}
for a detailed exposition. In the sequel we will especially  use  the
following  properties:
\begin{itemize}
 \item $\Res_X(f,g)$ belongs to the ideal $(f,g)\subset \SS[X],$
\item $\Res_X(f,g)$ is homogeneous of degree $n$, resp.~$m$, in the $a_i$'s, resp.~in the $b_j$'s,
\item $\Res_X(f,g)$ is homogeneous of degree $mn$ if we set $\deg(a_i):=m-i$ and $\deg(b_j):=n-j$ for all $i=0,\ldots,m$ and $j=0,\ldots,n$.
\end{itemize}

We also recall the definition of the principal subresultant of $f$ and $g$ that we will use later on. It is defined as the determinant of the above Sylvester matrix where the two last lines and columns number $n$ and $n+m$ are erased; more precisely
$$\SRes_X(f,g):=\left|
\begin{array}{ccccccccc}
  a_m & 0 & \cdots & 0 &  0 & b_n & 0 & 0 & 0 \\
  a_{m-1} & a_m &      & \vdots & \vdots & b_{n-1} & \ddots &  0 & 0   \\
  \vdots & a_{m-1} & \ddots &  0  &  0 & \vdots &  & b_n   & 0   \\
  a_{2} &  & \ddots & a_m & 0 & \vdots &  & b_{n-1} & b_n \\ 
   a_{1} &  a_{2} &  & a_{m-1} & a_m & b_{2} &  &   \vdots & b_{n-1} \\
a_0 & a_{1} & \ddots & \vdots & a_{m-1} & b_{1} &  &   \vdots & \vdots \\
 0 &  \ddots & \ddots  &  a_{2} & \vdots & b_{0} & \ddots & b_{2} & \vdots   \\
 \vdots&   & \ddots  & a_{1}  & a_{2} & 0 & \ddots & b_{1} & b_{2}   \\
   0 &  \cdots & 0  & a_0 & a_{1} & 0 & 0  &   b_0 & b_{1} \\
\end{array}\right|$$
(it is a square matrix of size $ (m+n-2)\times(m+n-2) $). Note that the subresultants share a lot of properties with the resultants; we refer the interested reader to \cite{ApJo}.

\subsubsection{The multivariate case}

All along this paper we will also use resultants of several homogeneous
polynomials; we now quickly recall this notion. Although they are usually defined
``geometrically'' (as equations of certain hypersurfaces obtained by
projection of an incidence variety), we will follow the formalism developed
by Jouanolou \cite{Jou91} because it easily provides many properties of
resultants.

Suppose given an integer $n\geq 1$, and a sequence of positive integers $d_1,\ldots,d_{n}$. One considers the $n$ ``generic'' homogeneous polynomials of degree $d_1,\ldots,d_{n}$, respectively, 
in the variables $X_1,\ldots,X_n$ (all assumed to have weight 1) :
$$f_i(X_1,\ldots,X_n)=\sum_{|\alpha|=d_i}U_{i,\alpha}X^\alpha, \ \ 
i=1,\ldots,n.$$
Denoting ${\UU}:=\ZZ[U_{i,\alpha}: i=1,\ldots,r, |\alpha|=d_i]$,  the polynomials $f_1,\ldots,f_n$ belongs to the ring ${C}:={\UU}[X_1,\ldots,X_n]$. 
The {\it ideal of inertia forms} of these polynomials is the ideal of ${C}$
  $$\mathrm{TF}_\mm(f_1,\ldots,f_n):=\{f\in {C} : \exists \nu
  \in \mathbb{N} \ \mm^\nu f \subset (f_1,\ldots,f_n)\} \subset {C}$$
where $\mm:=(X_1,\ldots,X_n)\subset C.$
It is naturally graded and it turns out that 
its degree zero graded part, denoted $\mathrm{TF}_\mm (f_{1},\ldots,f_{n})_0$, is a
principal ideal of $\UU$ and has a \emph{unique generator}, denoted
$\Res_{X_{1}:\cdots:X_{n}}$ or simply $\Res$,
which satisfies 
  \begin{equation}\label{normres}
 \Res(X_1^{d_1},\ldots,X_n^{d_n})=1.
 \end{equation}

To define the resultant of any given $n$-uples of homogeneous polynomials in the variables $X_1,\ldots,X_n$ (and also to clarify the left side of the equality \eqref{normres}) one proceeds as follows: 
Let $\SS$ be a commutative ring. For all integers $i\in
 \{1,\ldots,n\}$, suppose given a homogeneous polynomial of degree $d_i$ in the
    variables $X_1,\ldots,X_n$
    $$g_i=\sum_{|\alpha|=d_i}u_{i,\alpha}X^\alpha \in
    \SS[X_1,\ldots,X_n]_{d_i}$$
    and consider the morphism $\theta:{\UU}\rightarrow \SS:
    U_{j,\alpha} \mapsto u_{j,\alpha}$ which corresponds to the
    \emph{specialization} of the polynomials $f_i$ to the polynomials
    $g_i$.  Then, given an inertia form $a \in \mathrm{TF}_\mm(f_1,\ldots,f_n)$
    we set $a(g_1,\ldots,g_n):=\theta(a).$ In particular, the resultant of $g_1,\ldots,g_n$ is nothing but $$\Res(g_1,\ldots,g_n):=\theta(\Res(f_1,\ldots,f_n)).$$
Also, if $\SS={\UU}$ and $\theta$ is the identity (i.e.~$g_i=f_i$ for all $i$), then we get $a=a(f_1,\ldots,f_n)$; this clarifies the notation $\Res(f_1,\ldots,f_n)$ for the inertia form $\Res \in \UU$. 

\medskip

Resultants have a lot of interesting properties. We recall the ones we will use in the
sequel and refer the reader to \cite[\S5]{Jou91} for the proofs. Mention that many formulas are known to compute explicitly these resultants
 (e.g.~\cite{Mac02}, \cite{Jou97}, \cite{GKZ}, \cite{CLO} and the reference therein).

Let $\SS$ be any commutative ring and suppose given $f_1,\ldots,f_n$ homogeneous polynomials in the polynomial ring $\SS[X_1,X_2,\ldots,X_n]$ of positive degree $d_1,\ldots,d_n$ respectively.
\begin{itemize}
\item \emph{homogeneity:} for all $i=1,\ldots,n$, $\Res(f_1,\ldots,f_n)$ is homogeneous w.r.t.~the coefficients of $f_i$ of degree $d_1\ldots d_n/d_i$, 
\item \emph{isobarity:} $\Res(f_1,\ldots,f_n)$ is isobaric of degree $d_1\ldots d_n$ by giving to each coefficient of the $f_i$'s the power of its corresponding monomial in the variable $X_n$, 
 \item \emph{permutation of variables:} $$\Res(f_{\sigma(1)},\ldots,f_{\sigma(n)})=(\mathcal{E}(\sigma))^{d_1\ldots d_n}\Res(f_1,\ldots,f_n)$$ for any permutation $\sigma$ of the set $\{1,\ldots,n\}$ ($\mathcal{E}(\sigma)$ denotes the signature of the permutation $\sigma$),
\item \emph{elementary transformations:} $$\Res(f_1,\ldots,f_i+\sum_{i\neq j}h_jf_j,\ldots,f_n)=\Res(f_1,\ldots,f_n),$$
\item \emph{multiplicativity:} $$\Res(f_1'f_1'',f_2,\ldots,f_n)=\Res(f_1',f_2,\ldots,f_n)\Res(f_1'',f_2,\ldots,f_n)$$
\item \emph{base change formula:} if $g_1,\ldots,g_n$ are homogeneous polynomials in $\SS[X]$ of the same positive degree $d$, then 
\begin{multline}
\Res(f_1(g_1,\ldots,g_n),\ldots,f_n(g_1,\ldots,g_n))= \\ \Res(g_1,\ldots,g_n)^{d_1\ldots d_n}\Res(f_1,\ldots,f_n)^{d^{n-1}},$$
 \end{multline}

\item \emph{divisibility:} if $g_1,\ldots,g_n$ are homogeneous polynomials in $\SS[X]$ such that for all $i=1,\ldots,n$ there exists an integer $\mu_i$ such that $f_i \in (g_1,\ldots,g_n)^{\mu_i}$, then $\Res(g_1,\ldots,g_n)^{\mu_1\ldots \mu_n}$ divides $\Res(f_1,\ldots,f_n)$ in $\SS$.
\end{itemize}
In the following, we are going to consider resultant computation for
eliminating a subset $X_{i_{1}},\ldots, X_{i_{k}}$ of the variables. Such a
resultant, obtained by considering the homogenization of the polynomials
with respect to this variable subset, will be denoted hereafter
$\Res_{X_{i_{1}}, \ldots, X_{i_{k}}}$. With this notation, for homogeneous polynomials 
$f_1,\ldots,f_n \in \SS[X]$, we have
$$
\Res_{X_{1}:\cdots:X_{n}}(f_{1},\ldots,f_{n})=\Res_{X_{2},\ldots,
X_{n}}(f_{1}(1,X_{2},\ldots,X_{n}), \ldots, f_{n}(1,X_{2},\ldots,X_{n})).
$$

\subsection{Discriminants}
Given a polynomial
\[ P (X) \assign a_0 X^n + a_1 X^{n - 1} + \cdots + a_{n - 1} X + a_n \in \SS
   [X] \]
where $n \geq 1$ and $\SS$ is any commutative ring, we recall that the discriminant
of $P$, denoted $\Disc_X (P)$, satisfies the equality
\begin{align}\label{a0disc=res}
 a_0 \Disc_{X} (P (X)) = \Res_X (P (X), \partial_{X} P
(X)) = \Res_X ( \partial_{X} P (X),
   P (X))
\end{align}
where $\partial_X$ stands for the derivative with respect to the variable $X$, i.e.~$\partial/\partial_X$.
Its properties follow immediately from the ones of the resultant.
In particular, its degree in the coefficients of $P$ is $2(d-1)$. Note that
in our notation the polynomial $P$ is seen as a polynomial of its expected
degree, here $n$, as we did for the resultant. Indeed, in
\eqref{a0disc=res} the considered resultants are in degrees $(n,n-1)$ and
$(n-1,n)$ respectively.  

\medskip

Given a homogeneous polynomial $P(X_1,X_2,X_3)=\sum_{|\alpha|=d}U_\alpha
X_1^{\alpha_1}X_2^{\alpha_2}X_3^{\alpha_3}$ of degree $d$,  
its discriminant is the polynomial in the universal ring of
coefficients $\UU:=\ZZ[U_{\alpha}; |\alpha|=d]$, denoted
$\Disc_{X_{1}:X_2:X_{3}} (P)$ or simply
 $\Disc (P)$,
which satisfies  
\begin{equation}\label{eq:discr2}
 d^{d^2 - 3 d + 3} \Disc_{X_{1}:X_{2}:X_{3}} (P) = \Res_{X_{1}:X_{2}:X_{3}} (\partial_1 P, \partial_2 P,
   \partial_3 P)
\end{equation}
where $\partial_i$ stands for $\partial/\partial_{X_i}$ for all $i$.
It is an irreducible polynomial in $\UU$ which is homogeneous of degree $3 (d
- 1)^2$ in the $U_{\alpha}$'s. 
Note that we also have the following equality in $\UU$:
\begin{multline*}
 \Res_{X_{1}:X_{2}:X_{3}} (\partial_1 P, \partial_2 P, P) = \Disc_{X_{1}:X_{2}:X_{3}} (P)
   \Disc_{X_{2}:X_{3}} (P (0, X_2, X_3)) \\ = \Disc_{X_{1}:X_{2}:X_{3}}(P) \Disc_{X_3}
   (P (0, 1, X_3)) . 
\end{multline*}

\medskip

Finally, we also recall that being given two homogeneous polynomials $P_1$ and $P_2$ in the variables $X_1,X_2,X_3$ of degree $d_1,d_2$ respectively, their discriminant is defined by the formula \cite{Krull}
\[ \Disc_{X_{1}:X_{2}:X_{3}} (P_1, P_2) = \frac{\Res_{X_{1}:X_{2}:X_{3}}
   (P_1, P_2, \partial_2 P_1 \partial_3 P_2 - \partial_2 P_2 \partial_3
   P_1)}{\Res_{X_{1}:X_{2}:X_{3}} (P_1, P_2, X_1)} \in \UU \]
where $\UU$ denotes the universal ring of coefficients of $P_1$ and $P_2$ over $\ZZ$. 
It is an irreducible polynomial in $\UU$ which is homogeneous with respect to the
coefficients of $P_1$ of degree $d_2 (2 (d_1 - 1) + d_2 - 1)$ and
homogeneous with respect to the coefficients $P_2$ of degree $d_1
(2 (d_2 - 1) + d_1 - 1)$.

\subsection{Notations and derivatives}

Let $\SS$ be a commutative ring and suppose given a polynomial $P(X) \in \SS[X]$. We denote by $\partial_X P$ the formal derivative of $P$ with respect to the variable $X$. More precisely, the map
$$ \partial_X : \SS[X] \rightarrow \SS[X] : P \mapsto \partial_X P$$
is $\SS$-linear and for all integers $k\geq 0$ we have $\partial_X(X^k)=kX^{k-1}$. For any integer $i\geq 1$ we define $\partial_X^{i}$ as the composition of $\partial_X$ with itself $i$ times; for instance $\partial_X^i(X^k)$ is either equal to $0$ if $k<i$ or either equal to $k!/(k-i)!X^{k-i}$ if $k\geq i$.

Now, suppose given an element $a\in \SS$ and consider the $\SS$-linear map
$$ \delta_a : \SS[X] \rightarrow \SS[X] : P \mapsto \delta_a P:=\frac{P(X)-P(a)}{X-a}.$$
For instance, for all integers $k\geq 1$ we have $\delta_a(X^k)=\sum_{j=0}^{k-1}a^{k-1-j}X^j$.
As above, for any integer $i\geq 1$ we define $\delta_a^{i}$ as the composition of $\delta_a$ with itself $i$ times. It follows that for all integers $n\geq 1$ 
\begin{align}\label{Taylor1}
P(X)=P(a)+(X-a)\delta_a(P)(a)+(X-a)^2\delta_a^2(P)(a)+\cdots+(X-a)^n\delta_a^n(P)(X).
\end{align}
Indeed, we have $P(X)=P(a)+(X-a)\delta_a(P)(X)$ by definition. Applying this formula to $\delta_a(P)$ we deduce \eqref{Taylor1} for $n=2$. The general case is obtained by a similar induction. 

Although $\SS$ is only assumed to be a commutative ring, we still have the expected formula
\begin{lemma} For $k \ge 1$, for any $P\in \SS[X]$ and $a\in \SS$, we have the equality  $k!\,\delta_a^k P(a)=\partial_X^kP(a)$ in $\SS$. 
\end{lemma}
%
%
\begin{proof} 
Notice that $\delta_{a}P(X)$ is of degree $\deg(P)-1$ in the variable $X$. Thus if
$d > \deg(P)$, we have $\delta^d_{a}(P)=0$. We deduce that for any polynomial $P \in \SS[X]$, 
$$
P(X)=P(a)+(X-a)\delta_a(P)(a)+\cdots+(X-a)^d\delta_a^{\deg(P)}(P)(a).
$$
By the binomial identity, for any $n \in \NN$, 
$X^n = \sum_{i=0}^{n} \binom{n}{k} a^{n-k} (X-a)^{k}$.
By identification of the coefficients in the basis $(X-a)^{k}$ of $\SS[X]$, we deduce
that for all $ k\leq n$, we have $\delta_a^{k}(X^n)(a)=\binom{n}{k} a^{n-k}$.
As we have $\partial_X^k(X^n) = \frac{n!}{(n-k)!} X^{n-k}$, we deduce that
for any $k, n \in \NN$, $k!\,\delta_a^{k}(X^n)(a) = \partial_{X}^{k}(X^n)(a)$. 
By linearity, we also have this identity for any polynomial $P\in \SS[X]$.
\end{proof}

As a consequence, if $\SS$ is assumed to be a $\mathbb{Q}$-algebra then \eqref{Taylor1} shows that for all integers $n\geq 1$
$$P(X)=P(a)+(X-a)\partial_XP(a)+\cdots+(X-a)^n\frac{\partial_X^n P(a)}{n!}+(X-a)^{n+1}\delta_a^{n+1}(P).$$
If $n$ is bigger than the degree of $P$, then we get 
 the Taylor expansion formula 
$$P(X)=P(a)+\sum_{n\geq 1} \frac{\partial_X^n P(a)}{n!}(X-a)^n.$$

We now fix a notation that we will use all along the paper. We will mainly manipulate polynomials in the three variables
$X_1,X_2$ and $X_3$ over a commutative ring $\SS$. Introducing a new indeterminate $X_4$, for any polynomial $P
 \in \SS [X_1, X_2, X_3]$ we set 
$$\delta_{3, 4} (P) : = \frac{P (X_1, X_2, X_3) - P (X_1,
X_2, X_4)}{X_3 - X_4} \in \SS [X_1, X_2, X_3, X_4]$$ so that we have $$
P(X_{1},X_{2},X_{4})= P(X_{1},X_{2},X_{3}) + (X_{4}-X_{3})\delta_{3,4}(P).
$$
Observe that $\delta_{3,4}(P)$ is nothing but $\delta_{X_4}(P) \in (\SS[X_1,X_2,X_4])[X_3]$ and therefore that all the above definitions and formulas can be expressed with this notation. Therefore, we have
\begin{align} \label{eq1}
\delta_{3,4}P &= \partial_{3}P(X_{3}) + (X_{4}-X_{3})\delta_{3,4}^{2}P,
\end{align}
\begin{multline} \label{eq2}
2\delta^{2}_{3,4}P =\partial^{2}_{3}P(X_{3}) +
2(X_{4}-X_{3})\delta_{3,4}^{3}P\\
  =\frac{2}{(X_4 - X_3)^2} \left(P (X_1,
     X_2, X_3) - P (X_1, X_2, X_4) - (X_4 - X_3) \partial_{3} P_1 (X_1, X_2,
X_3)\right).
\end{multline}
The above definition can be extended to any variable. We denote by
$\delta^{2}_{i,j}$ the corresponding operation, $X_{i}$ playing the role of
$X_{3}$, and $X_{j}$ the role of $X_{4}$.
If the indices $i,j$ are omitted, we implicitly refer to the variables
$X_{3}$ and $X_{4}$. Notice that $\delta_{3,4}P_{|X_{4}=X_{3}}=\partial_{3}P(X_3)$. Also,
by convention we set $\delta_{i,i} P=\partial_{i}P$.

We will need the two following properties: for any polynomials $L,Q$, we have
\begin{eqnarray}
\delta_{3,4}(L\,Q) & = & \delta_{3,4}(L) \, Q(X_{3})+ L(X_{4})\, \delta_{3,4}(Q),
\label{eq:p1}\\
\delta_{3,4}^{2}(L\,Q) & = & \delta_{3,4}^{2}(L) \, Q(X_{3}) + \delta_{3,4}(L) \, \partial_{3}(Q)(X_{3})+ L(X_{4})\, \delta_{3,4}^{2}(Q). \label{eq:p2}
\end{eqnarray}
To prove \eqref{eq:p1}, we remark that 
$$
L(X_{3})Q(X_{3})- L(X_{4})Q(X_{4}) = Q(X_{3})(L(X_{4}) -L(X_{3})) + L(X_{4})
(Q(X_{4})-Q(X_{3}))
$$
and divide by $(X_{4}-X_{3})$ to get the formula. To prove \eqref{eq:p2}, we
substitute \eqref{eq1} in the previous relation and obtain
\begin{multline*}
(X_{4}-X_{3}) \partial_{3}(LQ) + (X_{4}-X_{3})^{2} \delta_{3,4}^{2}(QL) \\
=(X_{4}-X_{3}) (\partial_{3}(L) Q(X_{3}) + L(X_{3})\, \partial_{3}(Q)) + (X_{4}-X_{3})^{2}
\delta_{3,4}^{2} (QL)\\
 =   Q(X_{3})\left( (X_{4}-X_{3}) \partial_{3}(L) + (X_{4}-X_{3})^{2}\delta_{3,4}^{2}(L) \right) +
L(X_{4})  \left( (X_{4}-X_{3})\, \partial_{3}(Q)\right. \\ \left. + (X_{4}-X_{3})^{2} \delta_{3,4}^{2}(Q)\right)
\end{multline*}
so that
\begin{multline*}
(X_{4}-X_{3})^{2}
\delta_{3,4}^{2} (QL) =  (X_{4}-X_{3})^{2}\delta_{3,4}^{2}(L) Q(X_{3}) \\
+(X_{4}-X_{3}) (L(X_{4})- L(X_{3}))\, \partial_{3}(Q) 
+(X_{4}-X_{3})^{2}\, L(X_{4})\,\delta_{3,4}^{2}(Q),
\end{multline*}
from which we deduce the relation \eqref{eq:p2}.

Finally, mention that if $\SS$ is assumed to be a $\mathbb{Q}$-algebra then we have
\begin{multline*}
P(X_{1},X_{2},X_{4}) = P(X_{1},X_{2},X_{3}) + (X_{4}-X_{3}) \partial_{3} P(X_{3})+ \cdots + \\
\frac{1}{(i-1)!}\, (X_{4}-X_{3})^{i-1} \partial_{3}^{i-1} P(X_{3}) +  (X_{4}-X_{3})^{i}
\delta_{3,4}^{i} P
\end{multline*}
or equivalently $\delta_{3,4}^{i} P =\sum_{k\ge i}
\frac{1}{k!}(X_{4}-X_{3})^{k-1}\partial_{3}^{k} P(X_{3}).$
Similarly, we have
\begin{equation}
\delta_{3,4}^{i} P_{|X_{4}=X_{3}}= \frac{1}{i!}\partial^{i}_{3}P.
\end{equation}

\medskip

 Notice that we will very often omit variables $X_1$ and $X_2$
  in the sequel, especially in the proofs to avoid to overload the computations and the text.  

\subsection{A Bertini's lemma}

The elimination of the variables $X_1,X_2,X_3,X_4$ between polynomials constructed from the polynomials
\begin{eqnarray*}
  P_k (X_1, X_2, X_3):= \sum_{0 \leq i,  j ; i + j \leq d_k} &
  U^{(k)}_{i, j} X_1^i X_2^j X_3^{d_k - i - j} \in \UU [X_1, X_2, X_3], 
\end{eqnarray*}
$k = 1,\ldots,r$,
where $\UU$ denotes the universal coefficients ring $\UU :=
\mathbbm{Z}[U^{(k)}_{i, j}; 0 \leq i,  j ; i + j \leq d_k, k = 1, \ldots,
r]$, yields polynomials in $\UU$ for which we are going to give irreducible factorizations. As in geometric applications, this computation is applied with
the coefficients $U^{(k)}_{i, j}$ replaced by generic polynomials $\sum_{|
\alpha | \leq d_k - i - j} a^{(k)}_{i, j, \alpha} \tmmathbf{x}^{\alpha}$ of
degree $d_k - i - j \geqslant 0$, where we consider the coefficients
$a^{(k)}_{i, j, \alpha}$ as indeterminates  where $\tmmathbf{x}$ denotes a set of indeterminates $(x_1,\ldots,x_t)$ with $t\geq 1$. Therefore, we will need several times the
following lemma:

\begin{lemma}
  Suppose that $R$ is an irreducible homogeneous polynomial
  of $\UU$. Then the polynomial $R (\tmmathbf{x}, \tmmathbf{a})$ obtained
  by substituting the coefficients $U^{(k)}_{i, j}$ by generic polynomials
  $\sum_{| \alpha | \leq d_k - i - j} a^{(k)}_{i, j, \alpha}
  \tmmathbf{x}^{\alpha}$ is irreducible in $\mathbbm{Z}[\tmmathbf{x},
  a^{(k)}_{i, j, \alpha}]$.\label{lemma-irred}
\end{lemma}

\begin{proof}
  Assume that $R$ is not irreducible and can be splitted into the product of
  two factors in $\mathbbm{Z}[\tmmathbf{x},
  a^{(k)}_{i, j, \alpha}]$:
  \[ R (a^{(k)}_{i, j, \alpha}, \tmmathbf{x}) = R_1 (a^{(k)}_{i, j, \alpha},
     \tmmathbf{x}) R_2 (a^{(k)}_{i, j, \alpha}, \tmmathbf{x}) . \]
  By sending $\tmmathbf{x}$ to {\tmstrong{{\tmstrong{$\tmmathbf{0}$}}}} we
  observe that $R (a^{(k)}_{i, j, \alpha}, \tmmathbf{0})$ is an irreducible
  polynomial in the $a^{(k)}_{i, j, \alpha}$ since $R$ is irreducible so that
  we only have to rename each coefficient $a^{(k)}_{i, j, (0, \ldots, 0)}$ by
  $U^{(k)}_{i, j}$. It follows that either $R_1 (a^{(k)}_{i, j, \alpha},
  \tmmathbf{0})$ or $R_2 (a^{(k)}_{i, j, \alpha}, \tmmathbf{0})$ must be an
  invertible element in $\mathbbm{Z}$. But since $R_1$ and $R_2$ are
  homogeneous in the coefficients $a^{(k)}_{i, j, \alpha}$ this implies that
  either $R_1$ or $R_2$ is an invertible element in $\mathbbm{Z}$.
\end{proof}


\section{Resultant of resultants}

In all this section, we suppose given four positive integers $d_1, d_2, d_3,
d_4$ and four homogeneous polynomials
\begin{align*}
  P_k (X_1, X_2, X_3) = \sum_{0 \leq i,  j ; i + j \leq d_k} 
  U^{(k)}_{i, j} X_1^i X_2^j X_3^{d_k - i - j} \in \UU [X_1, X_2, X_3],  k = 1,\ldots,4,
\end{align*}
where $\UU$ denotes the universal  ring of coefficients $$\UU :=
\mathbbm{Z}[U^{(k)}_{i, j}; 0 \leq i,  j ; i + j \leq d_k, k = 1, \ldots,
4].$$  
We denote by $X_4$ a new indeterminate. Our first result is the most general situation of an iterated resultant with
four different polynomials. 

\begin{theorem}
  \label{prop-resres}
Defining
  \begin{eqnarray*}
    R_{12} & \assign & \Res_{X_3} (P_1 (1, X_2, X_3), P_2 (1, X_2, X_3))
    \in \UU [X_2],\\
    R_{34} & \assign & \Res_{X_3} (P_3 (1, X_2, X_3), P_4 (1, X_2, X_3))
    \in \UU [X_2],
  \end{eqnarray*}
  we have the following equality in $\UU$:
\begin{multline*}
\Res_{X_2} (R_{12}, R_{34}) = \\  \Res_{X_{1}:\cdots:X_{4}}(P_1 (X_1, X_2, X_3), P_2
    (X_1, X_2, X_3), P_3 (X_1, X_2, X_4), P_4 (X_1, X_2, X_4)). 
\end{multline*}
Moreover, the above quantity is non-zero, irreducible and multi-homogeneous with respect to the set of
  coefficients $(U^{(1)}_{i, j})_{i, j}$, $(U^{(2)}_{i, j})_{i, j}$, $(U^{(3)}_{i,
  j})_{i, j}$, $(U^{(4)}_{i, j})_{i, j}$ of multi-degree $(d_2 d_3 d_4, d_1 d_3
  d_4, d_1 d_2 d_4, d_1 d_2 d_3) .$
\end{theorem}

\begin{proof} First of all, we observe that the iterated resultant $\Res_{X_2} (R_{12}, R_{34})$ and the resultant
$$\mathcal{R}:=\Res(P_1 (X_1, X_2, X_3), P_2
    (X_1, X_2, X_3), P_3 (X_1, X_2, X_4), P_4 (X_1, X_2, X_4))$$
are both non-zero polynomials, for they both specialize to the quantity $(- 1)^{d_1 d_2 d_3 d_4}$ if   the polynomials $P_i (X_1, X_2, X_3), i = 1, \ldots, 4,$ are specialized  to
  $X_1^{d_1}$, $X_3^{d_2}$, $X_2^{d_3}$ and $X_3^{d_4}$ respectively. Note also that 
  the statement about the multi-degree of $\mathcal{R}$ follows from the homogeneity property of resultants.

  To prove the irreducibility of $\mathcal{R}$, we proceed by induction on the
positive integer $d \assign d_1 + d_2 + d_3 + d_4 \geq 4$. For $d = 4$, $\mathcal{R}$ equals the
  determinant
  \begin{eqnarray*}
    \left| \begin{array}{cccc}
      U^{(1)}_{1, 0} & U^{(2)}_{1, 0} & U^{(3)}_{1, 0} & U^{(4)}_{1, 0}\\
      U^{(1)}_{0, 1} & U^{(2)}_{0, 1} & 0 & 0\\
      U^{(1)}_{0, 0} & U^{(2)}_{0, 0} & U^{(3)}_{0, 0} & U^{(4)}_{0, 0}\\
      0 & 0 & U^{(3)}_{0, 1} & U^{(4)}_{0, 1}
    \end{array} \right| \in & \UU & 
  \end{eqnarray*}
  which is checked to be irreducible. Thus, we assume that $\mathcal{R}$ is
  irreducible up to a given integer $p \geq 4$ and we will prove that $\Rc$ is
  irreducible if $d = p + 1$. 
  To do this, first observe that one of the integers $d_1, d_2, d_3, d_4$
  must be greater or equal to 2. We can assume that $d_1 \geq 2$ without
  loss of generality. Consider the specialization $\phi$ leaving invariant the
  polynomials $P_2, P_3$ and $P_4$ and sending the polynomial $P_1$ to the
  product $L_{1} Q_{1}$ where $L_{1}$ and $Q_{1}$ are both generic forms of respective
  degree 1 and $d_1 - 1 \geq 1$. Then, by multiplicativity of resultants we have the equality
  \[ \phi (\mathcal{R}) = \Res (L_{1}, P_2, P_3, P_4) \Res (Q_{1},
     P_2, P_3, P_4), \]
  whose right hand side is a product of two irreducible polynomials by our
  induction hypothesis. As the specialization $\phi$ is homogeneous (in terms
  of the coefficients of the $P_i$'s, $L_1$ and $Q_{1}$), the number of
  irreducible factors of $\mathcal{R}$ can not decrease under the
  specialization $\phi$ and we deduce that $\mathcal{R}$ is the product of
  two irreducible polynomials $\Rc_1$ and $\Rc_2$. But then, one of these two
  factors must depend on the coefficients of $P_1$, say $\Rc_1$, and therefore
  $\phi (\Rc_1)$ must depend on the coefficients of $L_{1}$ and $Q_{1}$. This implies
  that $\Rc_2$ is an invertible element in $\mathbbm{Z}$ and consequently that
  $\mathcal{R}$ is irreducible.
  
  It remains to prove the claimed equality. To do this, we
  rewrite $P_k (1, X_2, X_3)$, for all  $k = 1, \ldots, 4$, as
  \begin{eqnarray*}
    P_k (1, X_2, X_3) = \sum_{i = 0}^{d_k} \left( \sum_{j = 0}^{d_k - i}
    U^{(k)}_{i, j} X_2^j \right) X_3^i \in \UU [X_2, X_3], &  & 
  \end{eqnarray*}
  and we then easily see from well-known properties of the Sylvester resultant
  that
  \begin{itemizedot}
    \item $R_{12}$ is bi-homogeneous in the set of coefficients $(U^{(1)}_{i,
    j})$ and ($U^{(2)}_{i, j}$) of bi-degree $(d_2, d_1)$,
    
    \item $R_{12}$ is a polynomial $\tmop{in} \UU[X_2]$ of degree $d_1 d_2$,
    
    \item $R_{12} \in (P_1 (1, X_2, X_3), P_2 (1, X_2, X_3)) \subset \UU[X_2,
    X_3]$.
  \end{itemizedot}
  Of course, completely analogous results hold for $R_{34}$, in particular
$$R_{34} \in (P_3 (1, X_2, X_4), P_4 (1, X_2, X_4)) \subset \UU[X_2,
    X_4].$$
Again, $\Res_{X_2}(R_{12},R_{34}) \in (R_{12},R_{34}) \subset \UU[X_2]$ and we deduce that
\begin{multline*} 
\Res_{X_2}(R_{12},R_{34}) \in  \\ (P_1(1,X_2,X_3),P_2(1,X_2,X_3),P_3(1,X_2,X_4),P_4(1,X_2,X_4)) \subset \UU[X_2,X_3,X_4].
\end{multline*}
After homogenization
  with the variable $X_1$, it follows that there exists
  an integer $N$ such that
\begin{multline*}
X_1^N \Res_{X_2}(R_{12},R_{34}) \in  \\ (P_1 (X_1, X_2, X_3), P_2 (X_1, X_2, X_3),P_3(X_1,X_2,X_4),P_4(X_1,X_2,X_4))
\end{multline*}
in  $\UU[X_1,X_2,X_3,X_4]$ (notice that this does not imply directly that $\Res_{X_2} (R_{12}, R_{34})$ is an inertia form because $P_1 (X_1, X_2, X_3)$, $P_2
     (X_1, X_2, X_3)$, $P_3 (X_1, X_2, X_4)$ and  $P_4 (X_1, X_2, X_4)$ are not
generic polynomials). It 
implies by the divisibility property of resultants, that
$\Rc=\Res(P_1(X_3),P_2(X_3),P_3(X_4),P_4(X_4))$  divides  the quantity 
\begin{multline*}
\Res(X_1^N\Res_{X_2} (R_{12}, R_{34}),P_1(X_1,X_2,X_3),P_2(X_1,X_2,X_3),P_3(X_1,X_2,X_4))= \\ \Res_{X_2} (R_{12}, R_{34})^{d_1d_2d_3}\Res(X_1,P_1(X_3),P_2(X_3),P_3(X_4))^N. 
\end{multline*}
Since $\mathcal{R}$ is irreducible and since the second term in the right hand side
of the above product does not depend on the coefficients of $P_4$, we deduce
that $\mathcal{R}$ divides $\Res_{X_2} (R_{12}, R_{34})$. 
Now, from the degree properties of
  $R_{12}$ and $R_{13}$ we deduce that $\Res_{X_2} (R_{12}, R_{34})$ is,
  similarly to $\mathcal{R}$, multi-homogeneous with respect to the set of
  coefficients $(U^{(1)}_{i, j})_{i, j}, (U^{(2)}_{i, j})_{i, j}, (U^{(3)}_{i,
  j})_{i, j},$ $(U^{(4)}_{i, j})_{i, j}$ of multi-degree 
$$(d_2 d_3 d_4, d_1 d_3
  d_4, d_1 d_2 d_4, d_1 d_2 d_3).$$ This shows  that
  $\Res_{X_2} (R_{12}, R_{34})$ and $\mathcal{R}$ are equal up to multiplication by an
  invertible element in $\mathbbm{Z}$. To determine this invertible element, we
  take again the specialization sending $P_1$ to $X_1^{d_1}$, $P_2$ to
  $X_3^{d_2}$, $P_3$ to $X_2^{d_3}$, $P_4$ to $X_3^{d_4}$, and check
  that $\mathcal{R}$ specializes to $(- 1)^{d_1 d_2 d_3 d_4}$, as well as
  $\Res_{X_2} (R_{12}, R_{34}).$
\end{proof}

A specialization of this theorem gives the following result.
\begin{corollary}
  \label{resres1}\label{resres}Given four
  polynomials $f_k (\tmmathbf{x}, y, z), k = 1, \ldots, 4$, of the form
  \[ f_k (\tmmathbf{x}, y, z) = \sum_{| \alpha | + i + j \leqslant d_k}
     a^{(k)}_{\alpha,i, j} \tmmathbf{x}^{\alpha} y^i z^j \in \SS [\tmmathbf{x}]
     [y, z], \]
  where $\tmmathbf{x}$ denotes a set of variables $(x_1, \ldots, x_n)$ for
  some integer $n \geq 1$ and $\SS$ is any commutative ring, then the iterated
  resultant $\Res_y (\Res_z (f_1, f_2), \Res_z (f_3, f_4))
  \in \SS [\tmmathbf{x}]$ is of degree at most $d_1 d_2 d_3 d_4$ in
  $\tmmathbf{x}$ and we have
  \begin{multline*} \Res_y^{} (\Res_z (f_1, f_2), \Res_z (f_3, f_4)) = \\
     \Res_{y, z, z'} (f_1 (\tmmathbf{x}, y, z), f_2 (\tmmathbf{x}, y,
     z), f_3 (\tmmathbf{x}, y, z'), f_4 (\tmmathbf{x}, y, z')) \in \SS
     [\tmmathbf{x}] . \end{multline*}
  Moreover, if the polynomials $f_1, f_2, f_3$ and $f_4$ are sufficiently
  generic then this iterated resultant is irreducible and has exactly degree
  $d_1 d_2 d_3 d_4$ in $\tmmathbf{x}$.
\end{corollary}

\begin{proof}
  It is a corollary of Theorem \ref{prop-resres}. Indeed, the claimed
  equality is deduced from the equality given in Theorem \ref{prop-resres}
  by substituting $X_2 \tmop{by} y, X_3 \tmop{by} z$, by seeing $X_1$ as the
  homogenization variable of the variables $(X_2,X_3)$ and by specializing each
  coefficient $U^{(k)}_{i, j}$ to $\sum_{| \alpha | \leq d_k - i - j}
  a^{(k)}_{i, j, \alpha} \tmmathbf{x}^{\alpha}$. Moreover, since each
  coefficient $U^{(k)}_{i, j}$ is specialized to a polynomial in
  $\tmmathbf{x}$ of degree at most $d_k$ we deduce the claimed degree bound as
  a consequence of the isobarity formula for resultants.
  
  If the polynomials $f_1, f_2, f_3$ and $f_4$ are sufficiently generic then
  it is clear that the degree bound is reached: this and the irreducibility statement
  is a consequence of Lemma \ref{lemma-irred}.
\end{proof}

The formula proved in Theorem \ref{prop-resres} can be specialized to get the
factorization of an iterated resultant which may occur quite often in practical situations (implicitization of a rational surface, projection of the intersection of two surfaces in projective space,\ldots~e.g.~\cite{CaMa,bem-upoicagd-03}).
\begin{proposition}
  \label{prop-resP1}Assume that $d_1 \geq 2$ and set
  \begin{eqnarray*}
    R_{12} & \assign & \Res_{X_3} (P_1 (1, X_2, X_3), P_2 (1, X_2, X_3))
    \in \UU[X_{2}],\\
    R_{13} & \assign & \Res_{X_3} (P_1 (1, X_2, X_3), P_3 (1, X_2, X_3))
    \in \UU[X_{2}].
  \end{eqnarray*}
  Then, we have the equality
  \begin{multline*}
    \Res_{X_2} (R_{13}, R_{12})  =  \Res_{X_{1}:X_{2}:X_{3}} (P_1 ,
    P_2, P_3) \times\\
     \Res_{X_{1}:\cdots:X_{4}} (P_1( X_3), P_2 ( X_4), P_3 (
    X_3), \delta_{3, 4} P_1 (X_3, X_4))
  \end{multline*}
  where the right hand side is a product of two irreducible polynomials in
  $\UU$.
  
  Moreover, the iterated resultant $\Res_{X_2} (R_{13}, R_{12})$ is
  multi-homogeneous with respect to the set of coefficients $(U^{(1)}_{i,
  j})_{i, j}, (U^{(2)}_{i, j})_{i, j}, (U^{(3)}_{i, j})_{i, j}$ of
  degree $2 d_1 d_2 d_3, d_1^2 d_3$ and  $d_1^2 d_2$ respectively.
\end{proposition}

\begin{proof}
  The claimed equality is easily obtained using formal properties of
  resultants by specialization of the formula proved in Theorem
  \ref{prop-resres}: 
\begin{eqnarray*}
\lefteqn{\Res(P_1(X_3),P_3(X_3),P_1(X_4),P_2(X_4))} \\ 
&=&\Res (P_1 (X_3), P_2 (X_4), P_3 (X_3), P_{1}(X_{4}) )  \\
&=&\Res (P_1 (X_3), P_2 (X_4), P_3 (X_3), P_{1}(X_{4})-P_{1}(X_{3}))  \\
&=&\Res (P_1 (X_3), P_2 (X_4), P_3 (X_3), (X_{4}-X_{3})\delta_{3,4}(P_{1}))  \\
&=&\Res (P_1, P_2 , P_3)\Res(P_1 (X_3), P_2 (X_4), P_3 (X_3),\delta_{3,4}(P_{1})).
\end{eqnarray*}
The multi-degree computation and the irreducibility of $\Res
  (P_1, P_2, P_3)$ are known properties of resultants. The only point
  which requires a proof is the irreducibility of the factor (which is easily seen
  to be non-zero by a straightforward specialization)
  \[ \mathcal{D} \assign \Res (P_1 (X_3), P_2 (
     X_4), P_3 ( X_3), \delta_{3, 4} (P_1) (X_3, X_4)) . \]
  To do this, we proceed similarly to what we did in Theorem
  \ref{prop-resres}, that is, by induction on the integer $d \assign d_1
  + d_2 + d_3 \geq 4$. We can check by hand (or with a computer) that $\mathcal{D}$ is an
  irreducible polynomial in $\UU$ if $(d_1, d_2, d_3) = (2, 1, 1) .$ \ We thus
  assume that $\mathcal{D}$ is irreducible up to a given integer $p \geq 4$
  and we will prove that $\mathcal{D}$ is irreducible if $d = p + 1$. If
  $d_2 \geq 2$ (resp. $d_3 \geq 2$) then we can specialize $P_2$ (resp. $P_3$)
  as a product of a generic linear form and a generic form of degree $d_2 - 1
  \geq 1$ (resp. $d_3 - 1 \geq 1$) and conclude, exactly as
  we did in Theorem \ref{prop-resres}, that $\mathcal{D}$ is then
  irreducible. Otherwise, then $d_1 \geq 3$ and we specialize $P_1$ to the
  product of a generic linear form
  \[ L_{1} (X_1, X_2, X_3) : = a\,X_1 + b\,X_2 + c\,X_3 \]
  and a generic form $Q_1$ of degree $d_1 - 1 \geq 2$. We call $\phi$ the map
  corresponding to this specialization. By \eqref{eq:p1}, we have
  \[ \delta_{3,4} (L_1Q_1) (X_3, X_4) = c\,Q_1(X_3) + L_{1} (X_4)\, \delta_{3,4} (Q_1) (X_3, X_4) \]
  and we deduce after some manipulations on resultants that
  \begin{eqnarray*}
    \phi (\mathcal{D}) & = & \Res (L_{1} (X_3) Q_1 (X_3), P_2 (X_4), P_3
    (X_3), c Q_1 (X_3) + L_{1} (X_4) \delta_{3,4} (Q_1)  )\\
    & = &  \Res (Q_1 (X_3), P_2 (X_4), P_3 (X_3), L_{1} (X_4) \delta_{3,4} (Q_1)
    (X_3, X_4))
    \times\\
    &  & \Res (L_{1} (X_3), P_2 (X_4), P_3 (X_3), c Q_1 (X_3) + L_{1} (X_4) \delta_{3,4} (Q_1) (X_3, X_4)) \\
    & = & \Res (Q_1 (X_3), P_2 (X_4), P_3 (X_3), \delta_{3,4} (Q_1)
    (X_3, X_4))\times \\
     & & \Res (Q_1 (X_3), P_2 (X_4), P_3 (X_3), L_{1} (X_4) )\times \\
    &  & \Res (L_1 (X_3), P_2 (X_4), P_3 (X_3), (L_{1} (X_4)-L_1(X_3)) \delta_{3,4} (Q_1) + c Q_1(X_3))\\
    & = & \Res (Q_1 (X_3), P_2 (X_4), P_3 (X_3), \delta_{3,4} (Q_1)(X_3, X_4))\times \\
    &  & \Res (Q_1 (X_3), P_2 (X_4), P_3 (X_3), L_{1} (X_4) )\times  \\ 
& & \Res (L_1 (X_3), P_2 (X_4), P_3 (X_3), c Q_1(X_4))\\
    & = & c^{d_2 d_3} Res (Q_1 (X_3), P_2 (X_4), P_3 (X_3), \delta_{3,4} (Q_1)
    (X_3, X_4))
    \times\\
    &  & \Res (Q_1 (X_3), P_2 (X_4), P_3 (X_3), L_{1} (X_4) )\times \\ 
& & \Res (L_1 (X_3), P_2 (X_4), P_3 (X_3),  Q_1(X_4)).
  \end{eqnarray*}
  Either by our induction hypothesis or by Theorem \ref{prop-resres}, it
  turns out that the three resultants involved in the right hand side of the
  above computation are irreducible in $\UU$. So if $\mathcal{D}$ were reducible, say
  $\mathcal{D}=\mathcal{D}_1 \mathcal{D}_2$, then each factor should be
  homogeneous in the coefficients $(U_{i, j}^{(1)})_{i, j}$ and hence $\phi
  (\mathcal{D}_1)$ and $\phi (\mathcal{D}_2)$ should be homogeneous in the
  coefficients of $Q_1$ and $L_{1}$ of the same degree. But
  \begin{align*}
    \deg_{L_{1}, Q_1} (c) & =  (1, 0),\\
    \deg_{L_{1}, Q_1} \left( \Res (L_{1} (X_3), P_2 (X_4), P_3 (X_3), Q_1
    (X_4)) \right) & =  ((d_1 - 1) d_2 d_3, d_2 d_3),\\
    \deg_{L_{1}, Q_1} \left( \Res (Q_1 (X_3), P_2 (X_4), P_3 (X_3), L_{1}
    (X_4)) \right) & =  ((d_1 - 1) d_2 d_3, d_2 d_3),\\
    \deg_{L_{1}, Q_1} \left( \Res (Q_1 (X_3), P_2 (X_4), P_3 (X_3), \delta_{3,4}
    (Q_1) (X_3, X_4)) \right) & =  (0, 2 d_1 d_2 d_3 - 3 d_2 d_3),
  \end{align*}
  which implies that either $\mathcal{D}_1$ or $\mathcal{D}_2$ is an
  invertible element in $\mathbbm{Z}$. 
\end{proof}

As a consequence of this proposition, we get the following result which
improves Theorem 3.1 and Theorem 3.2 in \cite{Callum}.
\begin{corollary}
  Given three polynomials $f_k (\tmmathbf{x}, y, z), k = 1, \ldots, 3$, of the
  form
  \[ f_k (\tmmathbf{x}, y, z) = \sum_{| \alpha | + i + j \leqslant d_k}
     a^{(k)}_{\alpha, i, j} \tmmathbf{x}^{\alpha} y^i z^j \in \SS [\tmmathbf{x}]
     [y, z], \]
  where $\tmmathbf{x}$ denotes a set of variables $(x_1, \ldots, x_n)$ for
  some integer $n \geq 1$ and $\SS$ is any commutative ring, then the iterated
  resultant $\Res_y (\Res_z (f_1, f_2), \Res_z (f_1, f_3))
  \in \SS [\tmmathbf{x}]$ is of degree at most $d_1^2 d_2 d_3$ in $\tmmathbf{x}$
  and we have
  \begin{multline*}
    \Res_y (\Res_z (f_1, f_2), \Res_z (f_1, f_3))  =  (-
    1)^{d_1 d_2 d_3} \times \\ \Res_{y, z} (f_1 (\tmmathbf{x}, y, z), f_2
    (\tmmathbf{x}, y, z), f_3 (\tmmathbf{x}, y, z)) \times \\
     \Res_{y, z, z'} (f_1 (\tmmathbf{x}, y, z), f_2 (\tmmathbf{x},
    y, z), f_3 (\tmmathbf{x}, y, z'),^{} \delta_{z, z'} (f_1)) .
  \end{multline*}
  Moreover, if the polynomials $f_1, f_2, f_3$ are sufficiently generic then
  this iterated resultant has exactly degree $d_1^2 d_2 d_3$ in $\tmmathbf{x}$
  and both resultants on the right hand side of the above equality are
  distinct and irreducible.
\end{corollary}

This corollary can be interpreted geometrically as follows. For simplicity we
assume that $\tmmathbf{x}$ is a unique variable $x$ and that $f_1$, $f_2$ and
$f_3$ are three polynomial in $x,y,z$. The resultant
$R_{12}:=\Res_z(f_1,f_2)$ defines the projection of the intersection curve
between the two surfaces $\{f_1=0\}$ and $\{f_2=0\}$. Similarly,
$R_{13}:=\Res_z(f_1,f_3)$ defines the projection of the intersection curve
between the two surfaces $\{f_1=0\}$ and $\{f_3=0\}$. Then the roots of
$\Res_y(R_{12},R_{13})$ can be decomposed into two distinct sets: the set of
roots $x_0$ such that there exists $y_0$ and $z_0$ such that
$f_1(x_0,y_0,z_0)=f_2(x_0,y_0,z_0)=f_3(x_0,y_0,z_0)$, and the set of roots
$x_1$ such that there exist two distinct points $(x_1,y_1,z_1)$ and
$(x_1,y_1',z_1')$ such that $f_1(x_1,y_1,z_1)=f_2(x_1,y_1,z_1)$ and
$f_1(x_1,y_1',z_1')=f_3(x_1,y_1',z_1')$. The first set gives rise to the term
$\Res_{x,y,z}(f_1,f_2,f_3)$ in the factorization of the iterated resultant
$\Res_y(\Res_{12},\Res_{13})$, and the second set of roots corresponds to the
second factor. 

\begin{remark}\label{Resmn}
Before going further, it is a good point to emphasize the fact that all the
formulas presented in this paper are universal in the sens that they remain \emph{true for
any specialization} of the coefficients  of the given
polynomials. It should be noticed that this is true if and only if we
take the univariate \emph{resultants in their expecting degrees}. For instance, in
the above corollary, $\Res_z(f_1,f_2)$ denotes the resultant w.r.t.~the
variable $z$ of the polynomials $f_1,f_2$ which are seen as polynomials of
degree $d_1$ and $d_2$ respectively (even if their degree is actually
lower for a given specialization). If one does not take care of this point, then one looses the
universal property of these formulas as in \cite[\S 7]{Callum} where it is
observed that if $f_1:=y^2+z+x$, $f_2:=-y^2+z$ and $f_3:=y^2+z$ then
$\Res_{y,z}(f_1,f_2,f_3)=0$ (for there is a base point at infinity) but
$\Res_y(\Res_z(f_1,f_2),\Res_z(f_1,f_3))=4x^2\neq 0$, considering $f_1,f_2$ and
$f_3$ as polynomials in $z$ of degree 1 and not 2. In such a case, formulas
similar to the ones we proved above require the use of more sophisticated
resultants which can take into account some particular structure of
polynomial systems. For instance, in this example one can take into 
account the presence of the base point defined by the ideal $(y^2,w)$, 
where $w$ is the homogenizing variable, by considering the residual resultant
 \cite{BEM01,BusPhD} (or the multi-homogeneous resultant which is the same here): we have the decomposition
$$(f_1,f_2,f_3)=(y^2,w)\left(
\begin{array}{ccc}
1 & -1 & 1 \\
z+xw & z & z 
\end{array}\right)$$
and we can check that the residual resultant equals $4x^2$ (up to a sign).

\end{remark}

\section{Resultants of discriminants}

In this section, we will give the factorization of three iterated resultants
corresponding to two cases of a resultant of a discriminant and a resultant
and to the case of a resultant of two discriminants. As we will see, the
first case we are going to treat yields the two others by suitable
specializations. 

\subsection{Resultant of a discriminant and a resultant}

\begin{proposition}\label{resdisc-prop1} Assume that $d_1\geq 2$ and 
  set
  \begin{eqnarray*}
    D_1 & \assign & \Disc_{X_3} (P_1 (1, X_2, X_3)) \in \UU[X_{2}],\\
    R_{23} & \assign & \Res_{X_3} (P_2(1,X_2,X_3),P_3(1,X_2,X_3)) 
    \in \UU[X_{2}].
  \end{eqnarray*}
  Then the following equality holds in $\UU$:
  \begin{multline*}
    (U_{0, 0}^{(1)})^{d_2d_3} \Res_{X_2} (D_1, R_{23}) =  \\
     \Res_{X_{1}:\cdots:X_{4}} (P_1 (X_1, X_2, X_3), \partial_3 P_1 (X_1, X_2, X_3), P_2
    (X_1, X_2, X_4), P_3 (X_1,X_2,X_4) ) .
  \end{multline*}
  Moreover, the iterated resultant $ \Res_{X_2} (D_1, R_{23}) \in \UU$ is
  irreducible and 
  multi-homo\-ge\-neous with respect to the coefficients $(U^{(1)}_{i, j})_{i, j}$, $(U^{(2)}_{i, j})_{i, j}$
  and $(U^{(3)}_{i, j})_{i, j}$ of multi-degree $$(2d_2d_3(d_1-1),d_1(d_1-1)d_3,d_1(d_1-1)d_2).$$
\end{proposition}
\begin{proof} By \eqref{a0disc=res}, we easily get that
\begin{multline*}
 \mathcal{D}:=(U_{0, 0}^{(1)})^{d_2d_3}\Res_{X_2} (D_1, R_{23})=\Res_{X_3}(U_{0, 0}^{(1)}D_1,R_{23})=\\
\Res_{X_2}(\Res_{X_3}(P_1(1,X_2,X_3),\partial_3 P_1(1,X_2,X_3)),R_{23})
\end{multline*}
and hence, using the formula proved in Theorem \ref{prop-resres} where we
  specialize the polynomials $P_2, P_3$ and $P_4$ to the polynomials $\partial_{3}  P_1, P_2$ and $P_3$ respectively, we deduce that, in $\UU$,
  \begin{align*}
     \mathcal{D} &= \Res (P_1 (X_1, X_2, X_3), \partial_3P_1 (X_1, X_2, X_3), P_2 (X_1, X_2, X_4),P_3 (X_1, X_2, X_4)).
  \end{align*}
  The classical multi-degree formula for resultants gives the claimed result
  concerning the multi-degree of the iterated resultant $\Res_{X_2} (D_1,
  R_{23})$. Now, we proceed by induction on the integer $d:=d_1+d_2+d_3\geq 4$ (remember $d_1\geq 2$, $d_2\geq 1$ and $d_3\geq 1$) to prove the irreducibility of the iterated resultant $\Res_{X_3}(D_1,R_{23})$.

First, if $d=4$, that is to say $d_1=2$ and $d_2=d_3=1$, then we check by hand that $\mathcal{D}$ equals an irreducible polynomial in $\UU$ times $U_{0, 0}^{(1)}$. 

We now assume that $d\geq 5$. If $d_3\geq 2$ then we consider the
specialization $\phi$ which sends $P_3$ to the product of the generic linear
form $L_{3}$ and the generic homogeneous polynomial $Q_3$ of degree
$d_3-1\geq 1$ and leave $P_1$ and $P_2$ invariant. We get, using the
multiplicativity of resultants  
\begin{multline*}
\phi(\mathcal{D})=\Res(P_1 (X_3), \partial_3P_1 (X_3),  P_2(X_4), L_{3}(X_4)) \\\Res(P_1 (X_3), \partial_3P_1 (X_3),L_{1}(X_4),P_2(X_4),Q_3(X_4)).
\end{multline*}
Therefore, by the induction hypothesis $\phi(\mathcal{D})=(U_{0,
0}^{(1)})^{d_2d_3}R_1R_2$ where $R_1$ and $R_2$ are irreducible polynomials
such that $R_1$ depends only on the coefficients of $L_{3}$ and $R_2$ depends
only on the coefficients of $Q_3$. Since $\phi$ is a homogeneous
specialization, each irreducible factor of $\mathcal{D}$ which depends on
$P_3$ must depends on $L_{3}$ and $Q_3$, we deduce that $\mathcal{D}$ has
only one irreducible factor depending on $P_3$. Moreover, the irreducible
factors of $\mathcal{D}$ which do not depend on $P_3$ are left invariant by
$\phi$ so we deduce that $\mathcal{D}$ equals $(U_{0, 0}^{(1)})^{d_2d_3}$
times an irreducible factor and we are done. If $d_2\geq 2$ then we can argue
exactly in the same way.  

So it only remains to consider the case where $d_2=d_3=1$. Let $\psi$ be the
specialization  which sends $P_1$ to the product of the generic linear form
$L_{1}:=aX_1+bX_2+cX_3$ times the generic homogeneous polynomial $Q_{1}$ of
degree $d_1-1$ and leaves $P_2$ and $P_3$ invariant. By the basic properties of
resultants and our induction hypothesis we get: 
\begin{eqnarray*}
\psi(\mathcal{D}) &= &\Res (L_{1}(X_3)Q_{1} (X_3), c\,Q_{1}(X_3)+L_{1}(X_3)\partial_3(Q_{1}(X_3)), P_2 (X_4),P_3 (X_4)) \\
&=& \Res (L_{1}(X_3),c\,Q_{1}(X_3),P_2 (X_4),P_3 (X_4))  \times \\ 
& &\Res (Q_{1}(X_3),L_{1}(X_3)\partial_3(Q_{1}(X_3)),P_2 (X_4),P_3 (X_4))\\
&=& (-1)^{d_1-1}c\,\Res(L_{1}(X_3),Q_{1}(X_3),P_2 (X_4),P_3(X_4))^2 \times \\
& &\Res(Q_{1}(X_3),\partial_3Q_{1}(X_3),P_2 (X_4),P_3 (X_4)) \\
& =& (-1)^{d_1-1}c(U_{0, 0}^{(1)})^{d_1-1} \Res(L_{1}(X_3),Q_{1}(X_3),P_2 (X_4),P_3(X_4))^2 \times R
\end{eqnarray*}
where $R$ is an irreducible polynomial which does not depend on the
coefficients of $L_{1}$ and does depend on the coefficients of $Q_{1},P_2$ of
$P_3$; in particular it has degree $2(d_1-2)$ in the coefficients of
$Q_{1}$. Observe that $\Res(L_{1}(X_3),Q_{1}(X_3),P_2 (X_4),P_3(X_4))$ is
irreducible by theorem \ref{prop-resres} and has degree $d_1-1$ in the
coefficients of $L_{1}$ and $1$ in the coefficients of $Q_{1}$. Since $\psi$
is an homogeneous specialization, each irreducible factor of $\mathcal{D}$
which depends on $P_1$ must depend on $L_{1}$ and $Q_{1}$ and have the same
degree with respect to the coefficients of these two polynomials. From this
property and the above computation we deduce that $\mathcal{D}$ has only one
irreducible factor that depends on $P_1$. Moreover, since the others
irreducible factors of $\mathcal{D}$ are left invariant by $\psi$ we deduce
that  $\mathcal{D}$ equals $(U_{0, 0}^{(1)})^{d_1}$ times an irreducible
polynomial. 
\end{proof}

\begin{corollary}
  Given three polynomials $f_k (\tmmathbf{x}, y, z), k = 1, 2,3$, of the form
  \[ f_k (\tmmathbf{x}, y, z) = \sum_{| \alpha | + i + j \leqslant d_k}
     a^{(k)}_{\alpha, i, j} \tmmathbf{x}^{\alpha} y^i z^j \in \SS [\tmmathbf{x}]
     [y, z], \]
  where $\tmmathbf{x}$ denotes a set of variables $(x_1, \ldots, x_n)$ for
  some integer $n \geq 1$ and $\SS$ is any commutative ring, then the iterated
  resultant $\Res_y (\Disc_z (f_1), \Res_z (f_2, f_3)) \in \SS
  [\tmmathbf{x}]$ is of degree at most $d_1(d_1 - 1)d_2d_3$ in $\tmmathbf{x}$
  and we have
  \begin{multline*}
    (a_{0, 0,d}^{(1)})^{d_2 d_3} \Res_y (\Disc_z (f_1), \Res_z
    (f_2, f_3))  = \\ \Res_{y, z} (f_1 (\tmmathbf{x}, y, z), \frac{\partial f_1}{\partial z} (\tmmathbf{x}, y,
    z),f_2(\tmmathbf{x}, y, z),f_3(\tmmathbf{x}, y, z)).
  \end{multline*}

  Moreover, if the polynomials $f_1, f_2, f_3$ are sufficiently generic then
  this iterated resultant has exactly degree $d_1(d_1 - 1)d_2d_3$ in
  $\tmmathbf{x}$ and the iterated resultant $\Res_y (\Disc_z (f_1), \Res_z
    (f_2, f_3))$  is irreducible.
\end{corollary}

We can now specialize the formula of Proposition \ref{resdisc-prop1} to get the factorization of two kinds of iterated resultants: the resultant of two discriminants and the resultant of a discriminant of a polynomial $f$ and a resultant of $f$ and another polynomial. We begin with the simplest one.

\subsection{Resultant of two discriminants of distinct polynomials}

\begin{proposition}\label{prop-irred1} Assume that $d_1\geq 2$ and $d_2\geq 2$ and 
  set
  \begin{eqnarray*}
    D_1 & \assign & \Disc_{X_3} (P_1 (1, X_2, X_3)) \in \UU[X_2],\\
    D_{2} & \assign & \Disc_{X_3} (P_2(1,X_2,X_3)) 
    \in \UU[X_2].
  \end{eqnarray*}
  Then the following equality holds in $\UU$:
  \begin{multline*}
    (U_{0, 0}^{(1)})^{d_2(d_2-1)} (U_{0, 0}^{(2)})^{d_1(d_1-1)}\Res_{X_2} (D_1, D_{2}) =  \\
     \Res_{X_{1}:\cdots:X_{4}} (P_1 (X_1, X_2, X_3), \partial_3 P_1 (X_1, X_2, X_3), P_2
    (X_1, X_2, X_4), \partial_3 P_2 (X_1,X_2,X_4) ) .
  \end{multline*}
  Moreover, the resultant $ \Res_{X_2} (D_1, D_{2}) \in \UU$ is
  irreducible and 
  bi-homogeneous with respect to the coefficients $(U^{(1)}_{i, j})_{i, j}$
  and $(U^{(2)}_{i, j})_{i, j}$ of bi-degree $$(2d_2(d_1-1)(d_2-1),2d_1(d_1-1)(d_2-1)).$$
\end{proposition}

\begin{proof} We set $\mathcal{D}:=(U_{0, 0}^{(1)})^{d_2(d_2-1)} (U_{0, 0}^{(2)})^{d_1(d_1-1)}\Res_{X_2} (D_1, D_{2})$. By \eqref{a0disc=res}, we have 
\begin{multline*}
 (U_{0, 0}^{(2)})^{d_1(d_1-1)}\Res_{X_2} (D_1, D_{2})=\Res_{X_2}(D_1,U_{0, 0}^{(2)}D_2)= \\
\Res_{X_2}(D_1,\Res_{X_3}(P_2(1,X_2,X_3),\partial_3 P_2(1,X_2,X_3)))
\end{multline*}
and hence, using the formula proved in Proposition \ref{resdisc-prop1}, where we
  specialize the polynomial $P_3$ to the polynomial $\partial_3P_2$, we deduce that, in $\UU$, $\mathcal{D}$ is equal to 
$$   \Res_{X_{1}:\cdots:X_{4}}  (P_1 (X_1, X_2, X_3), \partial_3P_1 (X_1, X_2, X_3), P_2 (X_1, X_2, X_4),\partial_{3} P_2 (X_1, X_2, X_4)).$$
  The classical multi-degree formula for resultants gives the claimed result
  concerning the bi-degree of the iterated resultant $\Res_{X_2} (D_1,
  D_{2})$. To prove its irreducibility, we proceed by induction on the integer $d:=d_1+d_2\geq 4$.

First, if $d=4$, that is to say $d_1=d_2=2$, then we check by hand (or with a
computer) that $\mathcal{D}$ equals an irreducible polynomial in $\UU$ times
the factor $(U_{0, 0}^{(1)})^{2} (U_{0, 0}^{(2)})^{2}$. We now assume that
$d\geq 5$. Without loss of generality we can also assume that $d_1\geq d_2$
(since the problem is completely symmetric in $P_1$ and $P_2$) and hence that
$d_1\geq 3$. Consider the specialization $\phi$ which sends $P_1$ to the
product of the generic linear form $L_{1}:=aX_1+bX_2+cX_3$ times the generic
homogeneous polynomial $Q_{1}$ of degree $d_1-1$ and leave $P_2$
invariant. Using properties of resultants we get: 
\begin{eqnarray*}
\phi(\mathcal{D}) &=& \Res (L_{1}(X_3)Q_{1} (X_3), c\,Q_{1}(X_3)+L_{1}(X_3)\partial_3(Q_{1}(X_3)), P_2 (X_4),\partial_{3} P_2 (X_4)) \\
&= &\Res (L_{1}(X_3),c\,Q_{1}(X_3),P_2 (X_4),\partial_{3} P_2 (X_4))\times  \\ 
& & \Res (Q_{1}(X_3),L_{1}(X_3)\partial_3(Q_{1}(X_3)),P_2 (X_4),\partial_{3} P_2 (X_4))\\
&= & c^{d_2(d_2-1)}\Res(L_{1}(X_3),Q_{1}(X_3),P_2 (X_4),\partial_{3} P_2
(X_4))^2 \times \\
& & \Res(Q_{1}(X_3),\partial_3Q_{1}(X_3),P_2 (X_4),\partial_{3} P_2 (X_4)).
\end{eqnarray*}
Using our inductive hypothesis and Proposition \ref{resdisc-prop1} we deduce
that $$\phi(\mathcal{D})=(c\,U_{0, 0}^{(1)'})^{d_2(d_2-1)} (U_{0,
0}^{(2)})^{d_1(d_1-1)} R_1^2\times R_2$$ 
($U_{0, 0}^{(1)'}$ being the coefficient of $X_3^{d_1-1}$ in $Q_1$) where $R_1$ is an irreducible
polynomial of degree $d_2(d_2-1)(d_1-1)$ in the coefficients of $L_{1}$ and
$d_2(d_2-1)$ in the coefficients of $Q_{1}$, and $R_2$ is an irreducible
polynomial independent of the coefficients of $L_{1}$ and of degree
$2d_2(d_2-1)(d_1-2)$ in the coefficients of $Q_{1}$. Note that we already know that $(U_{0, 0}^{(1)})^{d_2(d_2-1)}(U_{0, 0}^{(2)})^{d_1(d_1-1)}$ is a factor of $\mathcal{D}$. Moreover, 
since $\phi$ is an
homogeneous specialization, each remaining irreducible factor of $\mathcal{D}$ which
depends on $P_1$ must depend on $L_{1}$ and $Q_{1}$ with the same degree, so
we deduce that $R_1^2R_2$ comes from the same irreducible factor of
$\mathcal{D}$ and we conclude that
$\mathcal{D}$ equals $(U_{0, 0}^{(1)})^{d_2(d_2-1)}(U_{0, 0}^{(2)})^{d_1(d_1-1)}$  times an irreducible polynomial in $\UU$.
\end{proof}

\begin{corollary}
  Given two polynomials $f_k (\tmmathbf{x}, y, z), k = 1, 2$, of the form
  \[ f_k (\tmmathbf{x}, y, z) = \sum_{| \alpha | + i + j \leqslant d_k}
     a^{(k)}_{\alpha, i, j} \tmmathbf{x}^{\alpha} y^i z^j \in \SS [\tmmathbf{x}]
     [y, z], \]
  where $d_1,d_2\geq2$, $\tmmathbf{x}$ denotes a set of variables $(x_1, \ldots, x_n)$ for
  some integer $n \geq 1$ and $\SS$ is a commutative ring, then the iterated
  resultant $\Res_y (\Disc_z (f_1), \Disc_z (f_2)) \in \SS
  [\tmmathbf{x}]$ is of degree at most $d_1(d_1 - 1)d_2(d_2-1)$ in $\tmmathbf{x}$
  and we have
  \begin{multline*}
     (a_{0, 0,d_{1}}^{(1)})^{d_2(d_2-1)} (a_{0, 0,d_{2}}^{(2)})^{d_1(d_1-1)}\Res_y (\Disc_z (f_1), \Disc_z
    (f_2))  = \\ \Res_{y, z} (f_1 (\tmmathbf{x}, y, z),  \frac{\partial f_1}{\partial z} (\tmmathbf{x}, y,
    z)), f_2 (\tmmathbf{x}, y, z),  \frac{\partial f_2}{\partial z} (\tmmathbf{x}, y,
    z))).
  \end{multline*}

  Moreover, if the polynomials $f_1, f_2, f_3$ are sufficiently generic then
  this iterated resultant has exactly degree $d_1(d_1-1) d_2 (d_1 - 1)$ in
  $\tmmathbf{x}$ and  is irreducible.
\end{corollary}

\subsection{Resultant of a discriminant and a resultant sharing one polynomial}
We now turn to the second specialization of Proposition
\ref{resdisc-prop1} which is a little more intricate than the previous
one. Note that this iterated resultant has also been studied in 
\cite[Theorem 3.3]{Callum}. In order to improve the previous analysis, we begin with two
technical results. We recall that we sometimes omit the variables $X_1,X_2$, i.e.~we note $P(X_3)$ instead of $P(X_1,X_2,X_3)$ to not overload the text. 

\begin{lemma}\label{dec-lemma} For $d_{1}\ge 2$, 
\begin{multline*}
 (U_{0, 0}^{(1)})^{d_1 d_2} \Res_{X_2} (D_1, R_{12}) = \Res_{X_{1}:X_{2}:X_{3}} (P_1, 
    \partial_{3} P_1 (X_3),P_2 (X_3))^2 \times \\
    \Res_{X_{1}:X_2:X_3:X_{4}} (P_1(X_{3}), \partial_3P_1(X_{3}),P_2 (X_4),\delta^{(2)}_{3, 4} P_1(X_{3},X_{4})).
\end{multline*}
\end{lemma}
\begin{proof} By the formula proved in Proposition \ref{resdisc-prop1} where we
  specialize the polynomial $P_2$ to the polynomial $P_1$ and the polynomial
  $P_3$ to the polynomial $P_2$, we obtain after some manipulations
\begin{eqnarray}\notag  (U_{0, 0}^{(1)})^{d_1 d_2}
  \Res_{X_2} (D_1, R_{12}) & = &
  \Res(P_1(X_3),\partial_3P_1(X_3),P_1(X_4),P_2(X_4))\\ \notag & = &
  \Res(P_1(X_3),\partial_3P_1(X_3),P_2(X_4),(X_4-X_3)\delta_{3,4} P_1))\\
  & = & \Res (P_1 (X_3), \partial_{3} P_1 (X_3),P_2 (X_3))
  \times\\ \notag & & \hfil \Res (P_1 (X_3), \partial_{3} P_1 (X_3),
  P_2 (X_4), \delta_{3, 4} (P_1)).
\end{eqnarray}
Using formula \eqref{eq1} we can push forward this computation and get
 \begin{eqnarray}\label{compRD}
 \notag    (U_{0, 0}^{(1)})^{d_1 d_2} \Res_{X_2} (D_1, R_{12}) & = &
 \Res (P_1 ( X_3), 
    \partial_{3} P_1 ( X_3),P_2 ( X_3)) \times\\ \notag 
    &  & \Res (P_1 ( X_3), \partial_{3}
    P_1 (X_3),  P_2 (X_4),(X_4 - X_3) \delta^{(2)}_{3, 4}
    P_1)\\ 
    & = & \Res (P_1 (X_3), 
    \partial_{3} P_1 (X_3),P_2 (X_3))^2 \times\\ \notag
    &  & \Res (P_1 (X_3), \partial_3P_1(X_3),P_2 (X_4),\delta^{(2)}_{3, 4} P_1). \notag
  \end{eqnarray}
\end{proof}
 
\begin{lemma}\label{calc-lemme} Suppose that $d_1\geq 3$. In $\UU$, we have the equality 
$$\Res_{X_{1}:\cdots:X_{4}} (P_1,  \partial_{3}
    P_1, \delta^{(2)}_{3,4}P_1,P_2 (X_4))=(U_{0, 0}^{(1)})^{d_1 d_2}\,\Tm(P_1,P_2)$$
where $\Tm(P_{1},P_{2})$ is an irreducible polynomial in $\UU$ of bi-degree
$$((3\,d_{1}-1)\,(d_{1} -2)\, d_{2},d_{1}(d_{1}-1)(d_{1}-2))$$ in the
coefficients of $(P_{1},P_{2})$.
\end{lemma}
\begin{proof} 
We will denote by $\mathcal{R}$ the above resultant. We first use a
geometric argument to justify that $\mathcal{R}$ is the product of a certain
power of the coefficient $U_{0, 0}^{(1)}$ and a certain power of an
irreducible polynomial that we will denote $\Tm(P_1,P_2) \in \UU$.

We rewrite the polynomial $P_1$ as
\begin{align*}
 P_1(X_3) &= U_{0, 0}^{(1)} X_3^{d_1} + c_1 X_3^{d_1 - 1} + \cdots + c_{d_1 - 2} X_3^2 + c_{d_1 -
     1} X_3 + c_{d_1}
\end{align*}
 where the $c_i$'s are homogeneous polynomials in $\UU[X_1, X_2]$ of degree
  $i$ respectively;  we have
  \begin{eqnarray*}
    \partial_3 P & = & d_1U_{0, 0}^{(1)} X_3^{d_1 - 1} + (d_1 - 1) c_1 X_3^{d_1 - 2}
    + \cdots + 2 c_{d_1 - 2} X_3 + c_{d_1 - 1} 
\end{eqnarray*}
and
$$
    \delta_{3,4}^{(2)} P_1  =  U_{0, 0}^{(1)} \left( \frac{(\sum_{i=0}^{d_1-1}X_3^iX_4^{d_1-1-i})-d_1X_3^{d_1}}{X_4-X_3} \right) + \cdots+ 
	c_{d_1-3}\left( X_4+2X_3 \right) + c_{d_1 - 2} .
$$
Embedding $\mathbbm{Z}$
  into the algebraic closure $\overline{\QQ}$ of $\QQ$, the variety defined by the
  equation $\mathcal{R}= 0$ is the projection of the incidence variety
\begin{multline*}
  \mathcal{W} \assign \{(x_1 : x_2 : x_3 : x_4) \times (u_{i, j}^{(k)}) \in
     \mathbbm{P}_{\overline{\QQ}}^3 \times \mathbbm{A}_{\overline{\QQ}}^N \text{ such that } \\  
	 P_1(x_3)=\partial_3P_1(x_3)=\delta^{(2)}_{3,4}P_1(x_3,x_4) = P_2(x_4) = 0\}
\end{multline*}
  (where $\mathbbm{A}_{\overline{\QQ}}^N$ denotes the affine space whose coordinates are the $N:=(d_1+2)(d_1+1)+(d_2+2)(d_2+1)$ indeterminate coefficients, over $\overline{\QQ}$) by the canonical
  projection on the second factor $$\pi_2 : \mathcal{W} \subset \mathbbm{P}^3
  \times \mathbbm{A}^N \rightarrow \mathbbm{A}^N.$$ 

Consider the canonical
  projection of $\mathcal{W}$ onto the first factor $\pi_1 : \mathcal{W} \rightarrow
  \mathbbm{P}^3$, which is surjective, and denote by $D$ the line in
  $\mathbb{P}^3$ which is defined by the equations $X_1=0$ and $X_2=0$. On
  the one hand, we observe that for all $x\in \mathbb{P}^3\setminus D$ the
  fiber $\pi_1^{- 1} (x)$ is a linear space of codimension 4 in
  $\mathbbm{A}^{N}$; therefore the algebraic closure of
  $\mathcal{W}_{|\mathbb{P}^3\setminus D}$ is an irreducible variety of
  dimension $N-1$ in $\mathbbm{P}^3 \times \mathbbm{A}^N$ whose projection by
  $\pi_2$ gives an irreducible hypersurface in $\mathbbm{A}^N$. We denote by
  $\Tm(P_1,P_2)$ a defining equation of this irreducible hypersurface. 
On the other hand, for all $x\in D$ the fiber
  $\pi_1^{-1}(x)$ is always included in the linear space of equation
  $U_{0,0}^{(1)}=0$ (it is actually exactly this linear space if
  $x=(0:0:1:0)$ and the linear space $U_{0,0}^{(1)}=U_{0,0}^{(2)}=0$
  otherwise). Consequently, since $\pi_{2}(\mathcal{W})=\{U_{0,0}^{(1)}=0\}\cup
  \pi_{2}(\overline{\mathcal{W}|\mathbb{P}^3\setminus D})$ we deduce that
  $\mathcal{R}$ is of the form
  $$\mathcal{R}=c(d_1,d_2)(U_{0,0}^{(1)})^{a(d_1,d_2)}\Tm(P_1,P_2)^{r(d_1,d_2)}
  \in \UU$$ where $c(d_1,d_2),a(d_1,d_2),r(d_1,d_2)$ are constants that we
  have to determine (note that $\Tm(P_1,P_2)$ can be chosen in $\UU$ because we know that the variety defined by $\mathcal{R}=0$ can be defined by an equation in $\UU$ since it is a resultant of polynomials in $\UU$).

Now, we will prove by induction on the integer $d:=d_1+d_2 \geq 4$ that
\begin{align}\label{Rindhyp}
\mathcal{R}=(U_{0,0}^{(1)})^{d_1d_2}\Tm(P_1,P_2) \in \UU
\end{align}
that is, for all integers $d_1,d_2$ such that $d_1\geq 3$ and $d_2\geq 1$, we have $c(d_1,d_2)=r(d_1,d_2)=1$ and $a(d_1,d_2)=d_1d_2$. 

First, by Lemma \ref{dec-lemma} and using
the computation \eqref{compRD}, we note that we always have $c(d_1,d_2) \geq
d_1d_2$.

We check by hand (or with a computer) that the induction hypothesis \eqref{Rindhyp} is true for $d=4$, i.e.~$d_1=3$ and $d_2=1$ and we assume that $d\geq 5$. 
If $d_2\geq 2$ we specialize $P_2$ to the product of two generic homogeneous
polynomials, say $P_2'$ and $P_2''$; then $\mathcal{R}(P_1,P_2)$ specializes,
by multiplicativity of the resultants, to
$(U_{0,0}^{(1)})^{d_1d_2}\mathcal{R}(P_1,P_2')\mathcal{R}(P_1,P_2'')$. Since
each irreducible factor of $\mathcal{R}(P_1,P_2)$ must depend on $P_2'$ and
$P_2''$ and since we already know that $c(d_1,d_2)\geq d_1d_2$, we deduce
that $\mathcal{R}$ satisfies \eqref{Rindhyp} for all couple $(d_1,d_2)$ such
that $d_2\geq 2$.  

We now turn to the case where $d_1\geq 4$ and $d_2=1$. Consider the homogeneous specialization $\phi$ which sends $P_1$ to $X_3Q_1$ where $Q_1$ is the generic homogeneous polynomial of degree $d_1-1$. Using the properties of resultants we get
\begin{eqnarray*}
 \phi(\mathcal{R}) & = & \Res(X_3Q_1(X_3),Q_1(X_3)+X_3\partial_3Q_1(X_3),\delta_{3,4}^{(2)}(X_3Q_1),P_2(X_4)) \\
&= & \Res(X_3,Q_1(X_3),\delta_{3,4}^{(2)}(X_3Q_1),P_2(X_4)) \times \\ 
& &\Res(Q_1(X_3),X_3\partial_3Q_1(X_3),\delta_{3,4}^{(2)}(X_3Q_1),P_2(X_4)) \\
& = & \Res(X_3,Q_1(X_3),\delta_{3,4}^{(2)}(X_3Q_1),P_2(X_4))^2 \times \\ 
& & \Res(Q_1(X_3),\partial_3Q_1(X_3),\delta_{3,4}^{(2)}(X_3Q_1),P_2(X_4)).
\end{eqnarray*}
And since, by \eqref{eq:p2}, $\delta_{3,4}^{(2)}(X_3Q_1)=\partial_3Q_1(X_3)+X_4\delta_{3,4}^{(2)}Q_1$, we deduce that
\begin{multline*}
 \phi(\mathcal{R}) 
=  \Res(X_3,Q_1(X_3),\delta_{3,4}^{(2)}(X_3Q_1), \\ P_2(X_4))^2 \Res(Q_1(X_3),\partial_3Q_1(X_3),X_4,P_2(X_4))\times \mathcal{R}(Q_1,P_2).
\end{multline*}
Moreover, \eqref{eq1} implies that $\delta_{3,4}^{(2)}(X_3Q_1)=\partial_3Q_1(X_3)+X_4\delta_{3,4}^{(2)}Q_1$ evaluated at $X_3=0$ is equal to $$\delta Q_1(0,X_4)=\frac{Q_1(X_4)-Q_1(0)}{X_4}.$$ So finally, we deduce that
\begin{multline*}
 \phi(\mathcal{R}) 
 = (U_{0,0}^{(1)})^{(d_1-1)}\Res(Q_1(X_1,X_2,0),\delta Q_1(X_1,X_2,0,X_4),P_2(X_1,X_2,X_4))^2 \\ \Res(Q_1(X_3),\partial_3Q_1(X_3),P_2(X_1,X_2,0))\Tm(Q_1,P_2).
\end{multline*}
Since $\Tm(Q_1,P_2)$ is irreducible (by our induction hypothesis), it follows
that $r(d_1,d_2)$ must equal 1 for  each irreducible factor of the above
specialization must appear to a power which is a multiple of
$r(d_1,d_2)$. Also, since $P_2(X_1,X_2,0)$ is a generic linear form in $X_1$
and $X_2$ it turns out that
$\Res(Q_1(X_3),\partial_3Q_1(X_3),P_2(X_1,X_2,0))$ is, up to a linear change
of coordinates, a discriminant of a univariate polynomial: it equals
$U_{0,0}^{(1)}$ times an irreducible polynomial in $\UU$. Moreover, it is
easy to see that $U_{0,0}^{(1)}$ does not divide $\Res(Q_1(X_1,X_2,0),\delta
Q_1(X_1,X_2,0,X_4),P_2(X_1,X_2,X_4))$ (for this resultant does not vanish
under this condition). Therefore, we deduce that $a(d_1,d_2)\leq d_1d_2$ and
then that $a(d_1,d_2)=d_1d_2$. Finally, the three resultants in the above
specialization formula are clearly primitive\footnote{the gcd of their coefficients is $1$} 
(either by the induction hypothesis for the last one or either because it stays primitive after the
change of coordinate induced by the linear polynomial $P_2$) and it follows
that $c(d_1,d_2)=1$.

The formula on the degree is obtained as a direct consequence of the known degree formula for the resultants of several homogeneous polynomials.
\end{proof}

\begin{proposition}\label{prop-resdisc} Suppose that $d_1\geq 2$.
  We set
  \begin{eqnarray*}
    D_1 & \assign & \Disc_{X_3} (P_1 (1, X_2, X_3)) \in \UU,\\
    R_{12} & \assign & \Res_{X_3} (P_1 (1, X_2, X_3), P_2 (1, X_2, X_3))
    \in \UU.
  \end{eqnarray*}
  Then the following equality holds in $\UU$:
  \begin{eqnarray*}
   \Res_{X_2} (D_1, R_{12}) & = & \Res_{X_{1}:X_{2}:X_{3}} (P_1 , P_2 , \partial_{3} P_1)^2 \, 
    \Tm(P_1,P_2)
  \end{eqnarray*}
  where the irreducible polynomial $\Tm(P_1,P_2)$ has been defined in Lemma \ref{calc-lemme} for $d_1\geq 3$; if $d_1=2$ we set $\Tm(P_1,P_2):=1 \in \UU.$

  Moreover, $\Res_{X_{1}:X_{2}:X_{3}} (P_1, P_2, \partial_{3} P_1) \in \UU$ is
  irreducible and the iterated resultant $\Res_{X_2} (D_1, R_{12})$ is
  bi-homogeneous w.r.t.~the coefficients $(U^{(1)}_{i, j})_{i, j}$
  and $(U^{(2)}_{i, j})_{i, j}$ of bi-degree $(3 d_1 d_2 (d_1^2 - 1), d_1^2
  (d_1 - 1))$.
\end{proposition}
\begin{proof} 
First, the classical multi-degree formula for resultants gives the claimed
result for the bi-degree of the iterated resultant $\Res_{X_2} (D_1, R_{12})$.
By Lemma \ref{dec-lemma}, we have 
 \begin{multline}
    (U_{0, 0}^{(1)})^{d_1 d_2} \Res_{X_2} (D_1, R_{12}) 
   = \Res (P_1 (X_3), 
    \partial_{3} P_1 (X_3),P_2 (X_3))^2 \times\\ 
    \Res (P_1 (X_3), \partial_3P_1(X_3),P_2 (X_4),\delta^{(2)}_{3, 4} P_1) 
  \end{multline}
and by Lemma \ref{calc-lemme}, we know that 
   $$\Res ( P_1 (X_3), \partial_{3}
    P_1 (X_3), P_2 (X_4),\delta^{(2)}_{X_3, X_4} P_1) =(U_{0, 0}^{(1)})^{d_1 d_2}\Tm(P_1,P_2),$$
where $\Tm(P_1,P_2)$ is an irreducible polynomial, which implies the claimed formula. 

To conclude the proof, it only remains to prove the irreducibility of the resultant 
  \[ \mathcal{R} \assign \Res_{X_{1}:X_{2}:X_{3}} (P_1, P_2, \partial_{3} P_1). \]
 We proceed as we already did several times: by induction on the
  integer $d \assign d_1 + d_2 \geq 3$. We check by hand that $\mathcal{R}$ is irreducible if $d_1=2$ and $d_2=1$ and suppose that $d\geq 4$. If $d_2\geq 2$ then one specializes $P_2$ to a product of two generic forms and we conclude using the multiplicativity property of the resultant. Otherwise, we have $d_1\geq 3$ and one specializes $P_1$ to $L_1Q_1$ where $Q_1$ is the generic homogeneous polynomial of degree $d_1-1$ and $L_1:=aX_1+bX_2+cX_3$ is the generic linear form; this sends $\mathcal{R}$ to
$$(-1)^{d_1d_2}c^{d_2}\Res(Q_1,P_2,\partial_3Q_1)\Res(Q_1,P_2,L_1)^2$$
where $\Res(Q_1,P_2,\partial_3Q_1)$ is irreducible by our induction
 hypothesis and also where $\Res(Q_1,P_2,L_1)$ is irreducible for it is the resultant of three generic polynomials.
Examining the degrees in the coefficients of $Q_1$ and $L_1$ of the above factors and using the fact that each irreducible factor of $\mathcal{R}$ must specializes to a polynomial having the same degree in the coefficients of $Q_1$ and $L_1$, we deduce that $\mathcal{R}$ is irreducible.
\end{proof}
 
A specialization of this proposition gives 
the following result which covers and precises \cite[Theorem 3.3]{Callum}.
\begin{corollary}
  Given two polynomials $f_k (\tmmathbf{x}, y, z), k = 1, 2$, of the form
  \[ f_k (\tmmathbf{x}, y, z) = \sum_{| \alpha | + i + j \leqslant d_k}
     a^{(k)}_{\alpha, i, j} \tmmathbf{x}^{\alpha} y^i z^j \in \SS [\tmmathbf{x}]
     [y, z], \]
  where $d_1,d_2\geq 2$, $\tmmathbf{x}$ denotes a set of variables $(x_1, \ldots, x_n)$ for
  some integer $n \geq 1$ and $\SS$ is any commutative ring, then the iterated
  resultant $$\Res_y (\Disc_z (f_1), \Res_z (f_1, f_2)) \in \SS
  [\tmmathbf{x}]$$ is of degree at most $d_1^2 d_2 (d_1 - 1)$ in $\tmmathbf{x}$
  and we have
\begin{multline*}
    \Res_y (\Disc_z (f_1), \Res_z(f_1, f_2)) 
=   \\ \Res_{y, z} (f_1 (\tmmathbf{x}, y, z), f_2   (\tmmathbf{x}, y, z), \frac{\partial f_1}{\partial z} (\tmmathbf{x}, y,
    z))^2 \, \Tm(f_1(\tmmathbf{x}, y, z),f_2(\tmmathbf{x}, y, z))
  \end{multline*}
where we recall that, in $\SS[x]$, we have $\Tm(f_1(\tmmathbf{x}, y, z),f_2(\tmmathbf{x}, y, z))=1$  if $d_1=2$ and otherwise
\begin{multline*}(a_{0, 0, d}^{(1)})^{d_1 d_2}\Tm(f_1(\tmmathbf{x}, y, z),f_2(\tmmathbf{x}, y, z))= \\ \Res_{y, z, z'} (f_1 (\tmmathbf{x}, y, z), f_2 (\tmmathbf{x},
    y, z), \frac{\partial f_1}{\partial z} (\tmmathbf{x}, y, z'),\delta^{(2)}_{z, z'} f_1 (\tmmathbf{x}, y, z, z'))\end{multline*}
with
  \[ \delta^{(2)}_{z, z'} f_1 (\tmmathbf{x}, y, z, z') \assign
     \frac{\delta_{z, z'} (f_1) (\tmmathbf{x}, y, z, z') - \frac{\partial
     f_1}{\partial z} (\tmmathbf{x}, y, z)}{z' - z} \in \SS [\tmmathbf{x}] [y,
     z, z'] . \]
  
  Moreover, if the polynomials $f_1, f_2, f_3$ are sufficiently generic then
  this iterated resultant has exactly degree $d_1^2 d_2 (d_1 - 1)$ in
  $\tmmathbf{x}$ and  both $$\Res_{y, z} (f_1 (\tmmathbf{x}, y,
  z), f_2 (\tmmathbf{x}, y, z), \frac{\partial f_1}{\partial z}
  (\tmmathbf{x}, y, z))$$ and $\Tm(f_1,f_2)$ are irreducible polynomials.
\end{corollary}


\section{Discriminant of a resultant}

In this section, we suppose given two positive integers $d_1, d_2 \geq 2$ and
  two homogeneous polynomials
  \begin{eqnarray*}
    P_k (X_1, X_2, X_3) \assign \sum_{0 \leq i,  j ; i + j \leq d_k} &
    U^{(k)}_{i, j} X_1^i X_2^j X_3^{d_k - i - j} \in \UU[X_1, X_2, X_3], & k =
    1, 2, \label{pol:dires}
  \end{eqnarray*}
where, as usual, $\UU$ denotes the universal  ring of coefficients $\mathbbm{Z}[U^{(k)}_{i, j}]$. We will hereafter focus on the factorization of a discriminant of a resultant. 

\begin{lemma}\label{lemme:JSRes} In $\UU[X_{2}]$, setting $X_1=1$, we have
\begin{equation*}
\partial_{2}\Res_{X_{3}}(P_{1},P_{2}) =
 (-1)^{d_{1}+d_{2}}\left|\begin{array}{cc}
\partial_2 P_1 & \partial_3 P_1 \\
\partial_2 P_2 & \partial_3 P_2 
\end{array}\right|
\SRes_{X_3}(P_{1},P_{2}) \text{ \rm modulo } (P_{1}, P_{2}). \label{eq:diffres}
\end{equation*}
\end{lemma}
\begin{proof} 
Introduce a new indeterminate $U$ and set
\begin{align} \notag
P_1(1,X_2,X_3+U) &= a_{d_1}X_3^{d_1}+a_{d_1-1}X_3^{d_1-1}+\cdots+a_{1}X_3+a_0, \\ \notag
P_2(1,X_2,X_3+U) &= b_{d_2}X_3^{d_2}+b_{d_2-1}X_3^{d_2-1}+\cdots+b_{1}X_3+b_0,
\end{align}
where the $a_i$'s and the $b_j$'s are polynomials in $\UU[X_2,U]$.
Expanding the resultant 
$$\Res_{X_3}(P_1,P_2)=\left|
\begin{array}{ccccccc}
  a_{d_1} & 0 & \cdots & 0 & b_{d_2} & 0 & 0 \\
  a_{d_1-1} & a_{d_1} &      & \vdots & b_{d_2-1} & \ddots &    0   \\
  \vdots &  & \ddots &  0  & \vdots &  & b_{d_2}      \\
  a_0 &  &  & a_{d_1} & b_{1} &  & b_{d_2-1}  \\ 
  0 & a_0  &  & a_{d_1-1} & b_{0} &  &   \vdots  \\
 \vdots&   & \ddots  & \vdots & 0 & \ddots & b_{1}    \\
   0 &  \cdots & 0  & a_0 & 0 & 0  &   b_0  \\
\end{array}\right|$$
with respect to its two last rows, we get
$$ 
\Res_{X_{3}}(P_{1},P_{2}) = (-1)^{d_1+d_2}
\left|\begin{array}{ccc}
a_0 & b_0   \\ 
a_1 & b_1  \\ 
\end{array}\right|\SRes_{X_3}(P_{1},P_{2}) 
+ a_{0}^{2} \Delta_{1} + a_{0} b_{0} \Delta_{2} + b_{0}^{2} \Delta_{3},
$$
where $\Delta_{i}\ (i=1,2,3)$ are polynomials in the  $a_{i}$'s and
$b_{j}$'s. Taking the derivative with respect to the variable $X_2$, we deduce that
\begin{equation}\label{eq:lemme}
\partial_2\Res_{X_{3}}(P_{1},P_{2}) = (-1)^{d_1+d_2}
\left|\begin{array}{ccc}
\partial_2a_0 & \partial_2b_0   \\ 
a_1 & b_1  \\ 
\end{array}\right|\SRes_{X_3}(P_{1},P_{2}) \text{ modulo }
(a_0,b_0).\end{equation}
Now, it is easy to check that we have
$$a_0=P_1(1,X_2,U), \ b_0=P_2(1,X_2,U), \ a_1=\partial_3P_1(1,X_2,U) \  \ b_1= \partial_3P_2(1,X_2,U).$$
Moreover, by invariance property (change of bases formula) 
\begin{multline*}
\Res_{X_3}(P_1(1,X_2,X_3+U),P_2(1,X_2,X_3+U)) = \\ \Res_{X_3}(P_1(1,X_2,X_3),P_2(1,X_2,X_3)) 
= \Res_U(P_1(1,X_2,U),P_2(1,X_2,U))
\end{multline*}
and the same is true for the subresultant $\SRes_{X_3}(P_1,P_2)$. Therefore we deduce that \eqref{eq:lemme} is no\-thing but the claimed equality by substituting $X_3$ by $U$.
\end{proof}

This lemma implies the following factorization.
\begin{proposition}\label{DiscRes:firstfactor} In $\UU$, we have the equality
\begin{multline*}
    \Disc_{X_2} (\Res_{X_3} (P_1 (1, X_2, X_3), P_2 (1, X_2,
    X_3))) = (-1)^{(d_2+1)(d_1d_2+d_1-1)} \times  \\ \Res_{X_{1}:X_{2}:X_{3}}(P_1,P_2,\SRes_{X_3} (P_1 , P_2)) \times \Disc_{X_{1}:X_{2}:X_{3}}(P_1, P_2).
\end{multline*}
  This iterated resultant is bi-homogeneous with respect to the sets of
coefficients $U^{(1)}_{i, j}$ and $U^{(2)}_{i, j}$ respectively of bi-degree
$(2 d_2 (d_1 d_2 - 1) ; 2 d_1 (d_1 d_2 - 1)).$
\end{proposition}

\begin{proof} Set $\Rc_0:=\Res(P_1,P_2,X_1)$, $J_1:=\left|\begin{array}{cc}
\partial_2 P_1 & \partial_3 P_1 \\
\partial_2 P_2 & \partial_3 P_2 
\end{array}\right|$ and recall that, by definition, we have $\Rc_0\,\Disc_{X_{1}:X_{2}:X_{3}}(P_1, P_2)=\Res(P_1,P_2,J_1)$. Setting 
$$\Dc:=\Disc_{X_2} (\Res_{X_3} (P_1 (1, X_2, X_3), P_2 (1, X_2,
    X_3))) \in \UU,$$ it follows from \eqref{a0disc=res}  that 
\begin{multline*}
\Rc:= \Rc_0\Dc= \\ \Res_{X_2}(\Res_{X_3}(P_1(1, X_2, X_3),P_2(1, X_2, X_3)),\partial_2\Res(P_1(1, X_2, X_3),P_2(1, X_2, X_3))).
\end{multline*}
Now, using Lemma \ref{lemme:JSRes} we deduce that $\Rc$ belongs to the ideal 
$$(P_1(1,X_2,X_3),P_2(1,X_2,X_3),J_1(1,X_2,X_3)\,\SRes_{X_3}(P_1(1,X_2,X_3),P_2(1,X_2,X_3)))$$
which gives, after homogenization by $X_1$, the existence of an integer $N\geq 1$ such that
$$X_1^N\Rc \in (P_1(X_1,X_2,X_3),P_2(X_1,X_2,X_3),J_1(X_1,X_2,X_3)\SRes_{X_3}(P_1(X_1,X_2,X_3))).$$
Therefore, using the divisibility property (and others) of the resultants we deduce that 
\begin{align*}
\Res(P_1,P_2,J_1\SRes_{X_3}(P_1,P_2)) 
&=\Res(P_1,P_2,J_1)\times \Res(P_1,P_2,\SRes_{X_3}(P_1,P_2)) \\ 
&= \Rc_0\times \Disc(P_1,P_2)\times \Res(P_1,P_2,\SRes_{X_3}(P_1,P_2)) \end{align*}
divides $\Res(P_1,P_2,X_1^N\Rc)=\Rc_0^N\Rc^{d_1d_2}=\Rc_0^{N+d_1d_2}\Dc^{d_1d_2}$ in $\UU$. We know that $\Rc_0$ and $\Disc(P_1,P_2)$ are irreducible polynomials in $\UU$; just by comparing their degree we see that $\Rc_0$ does not divide $\Disc(P_1,P_2)$. Also, just by looking to the defining matrix of the subresultant, we have
$$\SRes_{X_3}(X_1^{d_1}+X_2^{d_1-1}X_3,X_3^{d_2})=(-X_2)^{(d_1-1)(d_2-1)}$$
so that $\Rc_0(X_1^{d_1}+X_2^{d_1-1}X_3,X_3^{d_2})=\Res(X_1^{d_1}+X_2^{d_1-1}X_3,X_3^{d_2},X_1)=0$ in $\UU$ and also
\begin{multline*}
\Res(X_1^{d_1}+X_2^{d_1-1}X_3,X_3^{d_2},\SRes_{X_3}(X_1^{d_1}+X_2^{d_1-1}X_3,X_3^{d_2}))= \\ \Res(X_1^{d_1}+X_2^{d_1-1}X_3,X_3^{d_2},(-X_2)^{(d_1-1)(d_2-1)}) \\ =\Res(X_1,X_3,-X_2)^{d_1d_2(d_1-1)(d_2-1)}=1.
\end{multline*}
Since $\Rc_0$ is irreducible,  it does not divide $\Res(P_1,P_2,\SRes_{X_3}(P_1,P_2))$ and therefore 
$$\Disc(P_1,P_2)\, \Res(P_1,P_2,\SRes_{X_3}(P_1,P_2)) \text{ divides } \Dc \text{ in } \UU.$$
By known degree properties, $\Dc$ has degree $d_2(2d_1d_2-2)$ in the coefficients of the polynomial $P_1$ and that the product $$\Disc(P_1,P_2)\,\Res(P_1,P_2,\SRes_{X_3}(P_1,P_2))$$ has degree 
$$d_2(2(d_1-1)+d_2-1)+d_2(d_1-1)(d_2-1)+d_1d_2(d_2-1)=d_2(2d_1d_2-2)$$
(note that $\SRes_{X_3}(P_1,P_2)$ is an homogeneous polynomial in $X_1,X_2$ of degree $(d_1-1)(d_2-1)$ which is also homogeneous in the coefficients of $P_1$, resp.~$P_2$, of degree $d_2-1$, resp.~$d_1-1$) 
With a similar computation for the degree with respect to the coefficients of $P_2$, we deduce that $\Dc$ and the product $$\Disc(P_1,P_2)\Res(P_1,P_2,\SRes_{X_3}(P_1,P_2))$$ are equal in $\UU$ up to a non-zero element in $\ZZ$, that we denote $c(d_1,d_2)$. 

To finish the proof, it remains to determine $c(d_1,d_2) \in \UU$ using, as usual, a suitable specialization. We choose the specialization $\phi$ such that
$$\phi(P_1)=X_3^{d_1}+A{X_1}^{d_1-1}{X_2}+B{X_1}^{d_1} \text{ and } \phi(P_2)={X_1}^{d_2-1}X_3-{X_2}^{d_2}.$$ 
It is easy to compute
$$ \Res_{X_3}(X_3^{d_1}+A{X_2}+B,X_3-{X_2}^{d_2}) = (-1)^{d_1}({X_2}^{d_1d_2}+A{X_2}+B)$$
and hence to deduce that
\begin{multline*}
\Disc_{X_2}(\Res_{X_3}(\phi(P_1)(1,X_2,X_3),\phi(P_2)(1,X_2,X_3))) = \\ \Disc_{X_2}((-1)^{d_1}({X_2}^{d_1d_2}+A{X_2}+B))= \\
 \Res_{X_2}({X_2}^{d_1d_2}+A{X_2}+B,d_1d_2{X_2}^{d_1d_2-1}+A).
\end{multline*}
Denoting $\Dc$ the above quantity, we have
\begin{align*}
(d_1d_2)^{d_1d_2-1}\Dc &= \Res_{X_{2}}(d_1d_2{X_2}^{d_1d_2}+d_1d_2A{X_2}+d_1d_2B,d_1d_2{X_2}^{d_1d_2-1}+A) \\
&= \Res_{X_2}((d_1d_2-1)AX_2+d_1d_2B,d_1d_2{X_2}^{d_1d_2-1}+A)\end{align*}
and finally
\begin{multline*}
(d_1d_2)^{d_1d_2-1}\Dc= \\ (-1)^{d_1d_2-1}(d_1d_2)^{d_1d_2-1}\left|
\begin{array}{lllll}
(d_1d_2-1)A & 0 & \cdots & 0 & d_1d_2 \\ 
d_1d_2B & (d_1d_2-1)A &  & \ldots & 0 \\ 
0 & d_1d_2B & \vdots & 0 & \ldots \\ 
\vdots &  & \vdots & (d_1d_2-1)A & 0 \\ 
0 & \ldots & 0 & d_1d_2B & A
\end{array}\right|,
\end{multline*}
that is to say,
$$\Dc=(-1)^{d_1d_2-1}(d_1d_2-1)^{d_1d_2-1}U^{d_1d_2}+(d_1d_2)^{d_1d_2}B^{d_1d_2-1}.$$
Now, since two consecutive integers are always relatively prime, we deduce
that $c(d_1,d_2)=\pm 1 \in \ZZ$. To determine exactly this integer, we
compute the other side of the claimed equality. For simplicity, we will
consider the specialization $\psi$ which is similar to the specialization
$\phi$ with in addition $A=0$. We have 
\begin{eqnarray*}
\lefteqn{\Disc(X_3^{d_1}+B{X_1}^{d_1},{X_1}^{d_2-1}X_3-{X_2}^{d_2})} & & \\
& = &\frac{\Res(X_3^{d_1}+B{X_1}^{d_1},{X_1}^{d_2-1}X_3-{X_2}^{d_2},d_1d_2{X_2}^{d_2-1}{X_3}^{d_1-1})}{\Res(X_3^{d_1}+B{X_1}^{d_1},{X_1}^{d_2-1}X_3-{X_2}^{d_2},X_1)} \\ 
&=&(-1)^{d_1(d_2+1)} (d_1d_2)^{d_1d_2}\Res(X_3^{d_1}+BX_1^{d_1},X_1^{d_2-1}X_3)^{d_2-1}\Res(BX_1^{d_1},-X_2^{d_2})^{d_1-1} \\
&=&(-1)^{d_1(d_2+1)} (d_1d_2)^{d_1d_2} (-1)^{d_1(d_2-1)^2}B^{d_2-1}B^{d_2(d_1-1)} \\
&=&(-1)^{d_1(d_2+1)+d_1(d_2-1)^2}(d_1d_2)^{d_1d_2}B^{d_1d_2-1}.
\end{eqnarray*}
It is not hard to see from the definition of the subresultant that
$$\SRes_{X_3}(X_3^{d_1}+BX_1^{d_1},X_1^{d_2-1}X_3-X_2^{d_2})=X_1^{(d_1-1)(d_2-1)}$$
and also to compute
$$\Res_{X_{1}:X_{2}:X_{3}}(X_3^{d_1}+BX_1^{d_1},X_1^{d_2-1}X_3-X_2^{d_2},X_1^{(d_1-1)(d_2-1)})=(-1)^{(d_1-1)(d_2-1)}.$$
Gathering all these specializations, we obtain
$c(d_1,d_2)=(-1)^{(d_2+1)(d_1d_2+d_1-1)}$, as claimed. 
\end{proof}

At this point, the factorization given in the above proposition is not
complete since we only know that one factor is irreducible. The following
result shows that the second factor is not irreducible, but is the square of
an irreducible polynomial and moreover that it can be interpreted
as a particular iterated resultant itself. 
\begin{lemma}\label{3eqs} Introducing a new indeterminate $X_4$, we have the
following equalities in $\UU$: 
\begin{multline*}
\Res_{X_{1}:X_{2}:X_{3}}(P_1,P_2,\SRes_{X_{3}}(P_1,P_2))=  \\
  \Res_{X_{2}}(\Res_{X_{3}}(P_1(1,X_2,X_3),P_2(1,X_2,X_3)),\SRes_{X_{3}}(P_1(1,X_2,X_3),P_2(1,X_2,X_3))) \\ 
=\Res_{X_{1}:X_2:X_3:X_{4}}(P_1,\delta_{3,4} P_1, P_2,\delta_{3,4} P_2).
\end{multline*}
\end{lemma}
\begin{proof} First, for simplicity we set 
\begin{multline*}
\Rc_1:= \Res_{X_{2},X_{3}}(\Res_{X_{3}}(P_1(1,X_2,X_3),P_2(1,X_2,X_3)), \\ \SRes_{X_3}(P_1(1,X_2,X_3),P_2(1,X_2,X_3))).
\end{multline*}
We know that $\Res_{X_3}(P_1,P_2)$ has degree $d_1d_2$ in $X_2$ and bi-degree
$(d_2,d_1)$ in $\UU$, and that $\SRes_{X_3}(P_1,P_2)$ has degree
$(d_1-1)(d_2-1)$ in $X_2$ and bi-degree $(d_2-1,d_1-1)$ in $\UU$. It follows
that $\Rc_1$ has bi-degree $((2 d_1 - 1) d_2 (d_2 - 1), (2 d_2 - 1) d_1 (d_1 - 1))$
which is exactly the same than the bi-degree of
$\Res_{X_{1}:X_{2}:X_{3}}(P_1,P_2,\SRes_{X_{3}}(P_1,P_2))$ 
(by a straightforward computation). Moreover, we have 
$$
\Rc_1 \in (P_1(1,X_2,X_3),P_2(1,X_2,X_3),\SRes_{X_{3}}(P_1(1,X_2,X_3),P_2(1,X_2,X_3)))
$$
which implies, after homogenization by $X_1$ and a suitable use of the divisibility property of the resultants that
$$
\Res_{X_{1}:X_{2}:X_{3}}(P_1,P_2,\SRes_{X_{3}}(P_1,P_2)) \text{ divides } \Res_{X_{1}:X_{2}:X_{3}}(P_1,P_2,X_1)^{N}\Rc_1^{d_1d_2}$$
in $\UU$, where $N$ denotes a positive integer. We have already seen in the
proof of Proposition \ref{DiscRes:firstfactor}, that
$\Rc_0:=\Res(P_1,P_2,X_1)$ is irreducible and does not divide
$\Res_{X_{1}:X_{2}:X_{3}}(P_1,P_2,\SRes_{X_{3}}(P_1,P_2))$. It follows that the later divides $\Rc$ and
since they have the same bi-degree in $\UU$ we deduce that they equal up to
multiplication by a constant. To determine this constant, we consider the
specialization $P_1=X_1^{d_1}+X_2^{d_1-1}X_3$ and $P_2=X_3^{d_2}$ for which
we have already seen that $\Res_{X_{1}:X_{2}:X_{3}}(P_1,P_2,\SRes_{X_{3}}(P_1,P_2))=1$ in the proof of
Proposition \ref{DiscRes:firstfactor}. Since we also have 
\begin{align*}
 \Rc_1(P_1,P_2) & = \Res_{X_2}((-X_1)^{d_1d_2},(-X_2)^{(d_1-1)(d_2-1)})=1.
\end{align*}
 we deduce that $\Rc_1=\Res_{X_{1}:X_{2}:X_{3}}(P_1,P_2,\SRes_{X_{3}}(P_1,P_2))$ in $\UU$.

We now turn to the proof of the third claimed equality. Introduce a new indeterminate $U$ and define
\begin{align} \notag
P_1(X_1,X_2,X_3+U) &= a_{d_1}X_3^{d_1}+a_{d_1-1}X_3^{d_1-1}+\cdots+a_{1}X_3+a_0, \\ \notag
P_2(X_1,X_2,X_3+U) &= b_{d_2}X_3^{d_2}+b_{d_2-1}X_3^{d_2-1}+\cdots+b_{1}X_3+b_0,
\end{align}
where the $a_i$'s and the $b_j$'s are polynomials in $\UU[X_1,X_2,U]$. The
subresultant $\SRes_{X_{3}}(P_1,P_2)$ is defined as the determinant of the matrix
$$\left(
\begin{array}{ccccccccc}
  a_{d_1} & 0 & \cdots & 0 &  0 & b_{d_2} & 0 & 0 & 0 \\
  a_{d_1-1} & a_{d_1} &      & \vdots & \vdots & b_{d_2-1} & \ddots &  0 & 0   \\
  \vdots & a_{d_1-1} & \ddots &  0  &  0 & \vdots &  & b_{d_2}   & 0   \\
  a_{2} &  & \ddots & a_{d_1} & 0 & \vdots &  & b_{d_2-1} & b_{d_2} \\ 
   a_{1} &  a_{2} &  & a_{d_1-1} & a_{d_1} & b_{2} &  &   \vdots & b_{d_2-1} \\
a_0 & a_{1} & \ddots & \vdots & a_{d_1-1} & b_{1} &  &   \vdots & \vdots \\
 0 &  \ddots & \ddots  &  a_{2} & \vdots & b_{0} & \ddots & b_{2} & \vdots   \\
 \vdots&   & \ddots  & a_{1}  & a_{2} & 0 & \ddots & b_{1} & b_{2}   \\
   0 &  \cdots & 0  & a_0 & a_{1} & 0 & 0  &   b_0 & b_{1} \\
\end{array}\right),$$
determinant which remains unchanged if we add, for all $i=1,\ldots,d_1+d_2-3$, the line number $i$ times $$\frac{X_3^{d_1+d_2-1-i} -X_4^{d_1+d_2-1-i} }{X_3-X_4}=\sum_{0 \leq i,j ; i+j=d_1+d_2-2-i} X_3^iX_4^j$$
to the last line which then becomes of the form
$$ \left(\begin{array}{cccccccccc} 
0& \cdots &0& a_0R_1& \delta_{3,4}(P_1)& 0 &\cdots &0 &b_0R_2 &\delta_{3,4}(P_2)
\end{array}\right)$$
where $R_1$ and $R_2$ are polynomials in $\UU[X_1,X_2,X_3,X_4,U]$. It follows that, by developping this determinant with respect to the last line, 
\begin{multline*}
\SRes_{X_{3}}(P_1(X_1,X_2,X_3+U),P_2(X_1,X_2,X_3+U))  \\ \in (a_0,\delta_{3,4}(P_1),b_0,\delta_{3,4}(P_2)) \subset
\UU[X_1,X_2,X_3,X_4].
\end{multline*}
Moreover, $a_0=P_1(X_1,X_2,U)$, $b_0=P_2(X_1,X_2,U)$ and by invariance of the subresultant under the change of coordinates $U \leftarrow X_3+U$ we deduce that 
\begin{multline*}
\SRes_{X_{3}}(P_1(X_1,X_2,U),P_2(X_1,X_2,U)) \\ \in (P_1(U),\delta_{3,4} P_1(U,X_4),P_2(U),\delta_{3,4} P_2(U,X_4))
\end{multline*}
which implies, after a substitution of $U$ by $X_3$ and a suitable use of the divisibility property of the resultants, that
\begin{multline*}
\Res_{X_{1}:\cdots:X_{4}}(P_1(X_3),\delta_{3,4} P_1(X_3,X_4),P_2(X_3),\delta_{3,4} P_2(X_3,X_4)) \\ \text{ divides } \Res_{X_{1}:X_{2}:X_{3}}(P_1,P_2,\SRes_{X_{3}}(P_1,P_2)).
\end{multline*}
An easy computation shows that these two resultants have the same bi-degree
w.r.t.~the coefficients of $P_1$ and $P_2$; therefore they are equal up to sign
in $\UU$ (we have already seen that 
$$\Res_{X_{1}:X_{2}:X_{3}}(P_1,P_2,\SRes_{X_{3}}(P_1,P_2))$$ is a
primitive polynomial in $\UU$ through a particular specialization). To
determine the sign we consider again the specialization
$P_1=X_1^{d_1}+X_2^{d_1-1}X_3$ and $P_2=X_3^{d_2}$ for which it is easy to
see that both resultants then specialize to 1. 
\end{proof}

The following result can be seen as the main explanation of 
\cite[theorem 3.4]{Callum}. 
\begin{proposition}
  \label{squareres} Introducing a new indeterminate $X_4$, there exists a non-zero irreducible polynomial $\Dm(P_{1},P_{2})$ in $\UU$ such that
$$
\Res_{X_{1}:\cdots:X_{4}} (P_1 , \delta_{3,4} (P_1) , P_2 , \delta_{3,4}(P_2))
= \Dm(P_{1},P_{2})^{2}.
$$
It is bi-homogeneous with respect to the set of coefficients $(U^{(1)}_{i,
j})_{i,j} \tmop{and} (U^{(2)}_{i, j})_{i, j}$ of bi-degree 
  \[ \text{$\left(\frac{(2 d_1 - 1) d_2 (d_2 - 1)}{2}, \frac{(2 d_2 - 1) d_1 (d_1 - 1)}{2}\right)$} . \]
\end{proposition}

\begin{proof} 
  Let us denote by $\mathfrak{R}$ the above resultant. Embedding $\mathbbm{Z}$
  into the algebraic closure of $\mathbbm{Q}$, the variety defined by the
  equation $\mathfrak{R}= 0$ is the projection of the incidence variety
\begin{multline*}  
\mathcal{W} \assign \\ \{(x_1 : x_2 : x_3 : x_4) \times (u_{i, j}^{(k)}) \in
     \mathbbm{P}^3 \times \mathbbm{A}^N \tmop{such} \tmop{that} P_1 = \delta_{3,4}
     (P_1) = P_2 = \delta_{3,4} (P_2) = 0\} 
\end{multline*}
  (where $N$ is the number of indeterminate coefficients) by the canonical
  projection on the second factor $\pi_2 : \mathcal{W} \subset \mathbbm{P}^3
  \times \mathbbm{A}^N \rightarrow \mathbbm{A}^N$. But for a generic point \
  $(a_{i, j}^{(k)}) \in \mathbbm{A}^N$ such that $\mathfrak{R}= 0$, we have at
  least two pre-image in $\mathcal{W}$ since if $(x_1 : x_2 : x_3 : x_4)$ is
  such pre-image then $(x_1 : x_2 : x_4 : x_3)$ is also a pre-image (which is
  generically different). It follows that the co-restriction of $\pi_2$ to the
  variety $\mathfrak{R}= 0$ has degree at least 2 and hence that, in $\UU$,
  \[ \mathfrak{R}= c (\tmmathbf{d}) \times \mathfrak{R}_1^{r_1 (\tmmathbf{d})}
     \times \mathfrak{R}_2^{r_2 (\tmmathbf{d})} \times \cdots \times
     \mathfrak{R}_p^{r_p (\tmmathbf{d})} \]
  where $p$ is a positive integer, $c (\tmmathbf{d})$ is also a positive integer
  but may depend on $\tmmathbf{d} \assign (d_1, d_2)$ and $r_1 (\tmmathbf{d}),
  \ldots, r_p (\tmmathbf{d})$ are positive integers greater or equal to 2 and
  may also depend on $\tmmathbf{d}$. Note that we know by Lemma \ref{3eqs} that $c (\tmmathbf{d})=1$. 
To determine the other quantities we will
  proceed by induction on $d_1 + d_2 \geq 4$ (remember that $d_1 \geq 2$ and $d_2
  \geq 2$). First, we can check by hand (or with a computer) that the claim is
  true if $d_1 = d_2 = 2$: $\mathfrak{R}$ is the square of an irreducible
  polynomial in $\UU$ of bi-degree (3,3). Now, assume that $d_1 + d_2 > 4$;
  without loss of generality one may assume that $d_1 \geq 3$. Consider the
  homogeneous specialization $\phi$ which sends $P_1$ to the product $L_{1} Q_1$
  where $Q_1$ is generic homogeneous polynomials of degree $d_1 - 1 \geq 2$
  and $L_{1} \assign \tmop{aX}_1 + \tmop{bX}_2 + \tmop{cX}_3$ is a generic linear
  form. We have, using obvious notations,
  \begin{eqnarray}\label{eq:dr1}
\\
    \phi(\mathfrak{R}) & = & \Res (L_{1} (X_3) Q_{1} (X_3), Q_{1} (X_3)
    \delta_{3,4} (L_{1}) + L_{1} (X_4) \delta_{3,4} (Q_{1}), P_2 (X_3), \delta_{3,4} (P_2)) \nonumber\\
    & = & \Res (L_{1} (X_3), c \, Q_1 (X_3) + L_{1} (X_4) \delta_{3,4} (Q_{1}),
    P_2 (X_3), \delta_{3,4} (P_2)) \times \nonumber\\
    &  & \Res (Q_{1} (X_3), L_{1} (X_4) \delta_{3,4} (Q_{1}), P_2 (X_3), \delta_{3,4}
    (P_2))\nonumber\\
    & = & \Res (L_{1} (X_3), c Q_{1} (X_3) + (L_{1} (X_4) - L_{1} (X_3))
    \delta_{3,4} (Q_{1}), P_2 (X_3), \delta_{3,4} (P_2)) \times \nonumber \\ 
    &  & \Res (Q_{1} (X_3), L_{1} (X_4), P_2 (X_3), \delta_{3,4} (P_2)) \times \nonumber \\
    & & \Res (Q_{1} (X_3), \delta_{3,4} (Q_{1}), P_2 (X_3), \delta_{3,4} (P_2))\nonumber\\
    & = & c^{d_2 (d_2 - 1)} \Res (L_{1} (X_3), Q_{1} (X_3) + (X_4 - X_3)
    \delta_{3,4} (Q_{1}), P_2 (X_3), \delta_{3,4} (P_2)) \times\nonumber\\
    &  & \Res (Q_{1} (X_3), L_{1} (X_4), P_2 (X_3), \delta_{3,4} (P_2)) \times \nonumber \\
    &  & \Res (Q_{1} (X_3), \delta_{3,4} (Q_{1}), P_2 (X_3), \delta_{3,4} (P_2))\nonumber\\
    & = & c^{d_2 (d_2 - 1)} \Res (L_{1} (X_3), Q_{1} (X_4), P_2 (X_3),
    \delta_{3,4} (P_2)) \times \nonumber \\ 
& & \Res (Q_{1} (X_3), L_{1} (X_4), P_2 (X_3), \delta_{3,4}
    (P_2)) \times\nonumber\\
    &  & \Res (Q_{1} (X_3), \delta_{3,4} (Q_{1}), P_2 (X_3), \delta_{3,4} (P_2))\nonumber\\
    & = & c^{d_2 (d_2 - 1)} \Res (L_{1} (X_3), Q_{1} (X_4), P_2 (X_3), \delta_{3,4} (P_2))^{2} 
\times\nonumber\\&&  \Res (Q_{1} (X_3), \delta_{3,4} (Q_{1}), P_2 (X_3), \delta_{3,4} (P_2)), \nonumber 
  \end{eqnarray}
since exchanging the role of $X_{3}$ and $X_{4}$, we have 
\begin{eqnarray*}
\lefteqn{\Res (L_{1} (X_3), Q_{1} (X_4), P_2 (X_3), \delta_{3,4} (P_2))}\\
& = & \Res (L_{1} (X_4), Q_{1} (X_3), P_{2}(X_{4}), \delta_{3,4} (P_2))\\
& = & \Res (L_{1} (X_4), Q_{1} (X_3), P_{2}(X_{3}), \delta_{3,4} (P_2))\\
& = & \Res (Q_{1} (X_3), L_{1} (X_4), P_{2}(X_{3}), \delta_{3,4} (P_2)).
\end{eqnarray*} 
  Observe that this resultant is irreducible by
  Proposition \ref{prop-resP1} and that the last one if the square of an
  irreducible polynomial by our induction hypothesis. Moreover,
  \begin{itemizedot}
    \item $c$ has bi-degree $(1, 0)$ in terms of the coefficients of $L_{1}$ and
    $Q_{1}$ respectively,
    
    \item $\Res (L_{1}(X_{4}), Q_{1} (X_3), P_2 (X_3), \delta_{3,4} (P_2))$ has also
    bi-degree $(d_2 (d_2 - 1) (d_1 - 1), d_2 (d_2 - 1))$ and
    
    \item $\Res (Q_{1} (X_3), \delta_{3,4} (Q_{1}), P_2 (X_3), \delta_{3,4} (P_2))$ is
    the square of an irreducible polynomial which has bi-degree $(0, d_2 (d_2
    - 1) (d_1 - 1) - \frac{d_2 (d_2 - 1)}{2})$.
  \end{itemizedot}
  Since the specialization $\phi$ is homogeneous, each irreducible factor
  $\mathfrak{R}_i$ of $\mathfrak{R}$ must give through this specialization
  irreducible polynomial(s) having the same degree with respect to the
  coefficients of $L_{1}$ and $Q_{1}$. With the bi-degree given above, we deduce
  that $\mathfrak{R}$ can at most have two irreducible factors (and moreover
  that they specialize to the same polynomial via $\phi$). It follows that we have $p
  = 1$ and $r_1 (\tmmathbf{d}) = 2$.
\end{proof}

\begin{remark} Some technical limitations of the theory of {\it anisotropic
resultants} as exposed in \cite{Jou91,Jou96} prevent the explicit description of the
``squareroot'' of the above resultant.  
More precisely, suppose given two sequences of integers $(m_1,\ldots,m_n)$ and $(d_1,\ldots,d_n)$ such that $m_j | d_i$ for all couple $(i,j)\in \{1,\ldots,n\}^2$. If $g_1,\ldots,g_n$ are $n$ homogeneous polynomials of degree $m_1,\ldots,m_n$ respectively in the graded ring $\SS[X_1,\ldots,X_n]$ (with $\deg(X_i)=1$ for all $i=1,\ldots,n$) and $f_1,\ldots,f_n$ are $n$ isobaric polynomials of weight $d_1,\ldots,d_n$ respectively in the graded ring ${}^a \SS[X_1,\ldots,X_n]$ (now with $\deg(X_i)=m_i$ for all $i=1,\ldots,n$) then an easy adaptation of the proof of the base change formula \cite[5.12]{Jou91} shows that
$$\Res(f_1(g_1,\ldots,g_n),\ldots,f_n(g_1,\ldots,g_n))= {}^a\Res(f_1,\ldots,f_n)^\Delta \Res(g_1,\ldots,g_n)^{\frac{d_1\ldots d_n}{m_1\ldots m_n}}$$
in $\SS$ where $\Delta:=\frac{m_1\ldots m_n}{\gcd(m_1,\ldots,m_n)}$. 

In our case, denoting $\delta^0(P):=P(X_1,X_2,X_3)-P(X_1,X_2,X_4)$, it is easy to see that
\begin{multline*}
\Res(\delta^0(P_1),\delta^1 (P_1),\delta^0(P_2),\delta^1(P_2))= 
2^{(d_1-1)(d_2-1)(d_1+d_2)} \times \\ \Res(P_1(X_1, X_2, X_3), \delta (P_1) (X_1, X_2, X_3,
    X_4), P_2 (X_1, X_2, X_3), \delta (P_2) (X_1, X_2, X_3, X_4)).
\end{multline*}
And since, for all $i=1,2$, $\delta^0(P_i)$ and $\delta^1(P_i)$ are symmetric polynomials with respect to the variables $X_3$ and $X_4$ we deduce that there exists four {\it quasi}-homogeneous polynomials $Q_i(X_1,\ldots,X_4)$, $i=1,\ldots,4$, with $\deg(X_1)=\deg(X_2)=\deg(X_3)=1$, $deg(X_4)=2$ and such that, for instance,
$$Q_1(X_1,X_2,X_3+X_4,X_3X_4)=\delta^0(P_1)(X_1,X_2,X_3,X_4).$$
By using the above adapted base change formula, we should obtain
$$\Res(\delta^0(P_1),\delta^1 (P_1),\delta^0(P_2),\delta^1(P_2))=\pm {}^a\Res(Q_1,Q_2,Q_3,Q_4)^2.$$
It turns out that $Q_1,Q_2,Q_3,Q_4$ are isobaric of weights
$(d_1,d_1-1,d_2,d_2-2)$ respectively and hence the condition of the existence
of their anisotropic resultant as in \cite[\S 2]{Jou96} are not fulfilled. However, in a personal communication Jouanolou informed us that it is possible to extend the theory of anisotropic resultant to our particular setting and conclude to the existence of such an anisotropic resultant. 
\end{remark}

Gathering the results of this section, we obtain the full factorization of
the discriminant of a resultant. 
\begin{theorem}
  \label{DiscRes}Introducing a new indeterminate $X_4$, we have in $\UU$:
\begin{multline*}\Disc_{X_2} (\Res_{X_3} (P_1 (1, X_2, X_3), P_2 (1, X_2,
    X_3)))= (-1)^{(d_2+1)(d_1d_2+d_1-1)}\times \\ \Disc_{X_{1}:X_{2}:X_{3}}(P_1,P_2)\Dm(P_1,P_2)^2\end{multline*}
 where  $\Disc_{X_{1}:X_{2}:X_{3}}(P_1,P_2)$ and $\Dm(P_1,P_2)$ are
irreducible polynomials in $\UU$. This iterated resultant is
  bi-homogeneous with respect to the sets of coefficients $U^{(1)}_{i, j}$ and
  $U^{(2)}_{i, j}$ respectively of bi-degree $$(2 d_2 (d_1 d_2 - 1) ; 2 d_1
  (d_1 d_2 - 1)) .$$
\end{theorem}

As usual, we can specialize this result to obtain the following:
\begin{corollary}
  \label{discres1}Given two polynomials $f_k (\tmmathbf{x}, y, z), k = 1, 2$,
  of the form
  \[ f_k (\tmmathbf{x}, y, z) = \sum_{| \alpha | + i + j \leqslant d_k}
     a^{(k)}_{\alpha,i, j} \tmmathbf{x}^{\alpha} y^i z^j \in \SS [\tmmathbf{x}]
     [y, z], \]
  where $\tmmathbf{x}$ denotes a set of variables $(x_1, \ldots, x_n)$ for
  some integer $n \geq 1$ and $\SS$ is any commutative ring, the iterated
  resultant $\Disc_y (\Res_z (f_1, f_2)) \in \SS [\tmmathbf{x}]$ is
  of degree at most $d_1 d_2 (d_1 d_2 - 1)$in $\tmmathbf{x}$ and can be factorized, up
  to sign, as
$$  \Disc_y (\Res_z (f_1, f_2))=(-1)^{(d_2+1)(d_1d_2+d_1-1)} \Disc_{y, z} (f_1,
    f_2)\Dm(f_1,f_2)^2.$$

  Moreover, if the polynomials $f_1$ and $f_2$ are sufficiently generic, then
  this iterated resultant has exactly degree $d_1 d_2 (d_1 d_2 - 1)$ in
  $\tmmathbf{x}$ and the two terms in the right hand side of the above
  equality are respectively an irreducible polynomial and the square of an irreducible polynomial.  
\end{corollary}

\section{Discriminant of a discriminant}
In this section, we are interested in analyzing two iterated discriminants. 
Before going into the algebraic study, let us consider the problem from a
geometric point of view. Suppose we are given an implicit surface
$f(x,y,z)=0$. Computing the discriminant of $f$ in $z$ consists in projecting
the apparent contour (or polar) curve in the $z$ direction, which is defined
by  $f(x,y,z)=0$, $\partial_{z} f(x,y,z)=0$. Computing the discriminant in
$y$ of this discriminant in $z$ of $f$ consists in computing the position of lines
parallel to the $y$ axis, which are tangent to the projected curve. 

We illustrate it by some explicit computations, with the polynomial
$$ 
f(x,y,z) = {z}^{4}-{y}^{3}z+2\,{z}^{3}-y{z}^{2}-{y}^{2}-x\,z+1.
$$
Its discriminant in $z$ is a polynomial of degree $12$ and the discriminant in $y$ of
this discriminant can be factorized as:
\begin{multline*}
5540271966595842048 \, 
(14348907\,{x}^{10}-93002175\,{x}^{9}+273574017\,{x}^{8}
- \\ 909290448\,{x}^{7}+ 
2868603336\,{x}^{6} -5353192260\,{x}^{5}
+9038030571\,{x}^{4}-17693165669\,{x}^{3}+ \\ 17648229264\,{x}^{2}-  4081683588\,x+218938829 ) ( x-1 ) ^{2} 
 (125\,x-173 ) ^{2}( 47832147
\,{x}^{4}+ \\ 147495688\,{x}^{3}-  245928792\,{x}^{2}-212731008\,x+230501936 )^{3}.
\end{multline*} 
Figures\footnote{The
topology computation and visualization have been performed by the softwares
{\sc axel} (\texttt{http://axel.inria.fr/}) and
{\sc synaps} (\texttt{http://synaps.inria.fr/}).} 1 and 2 illustrate the situation, where we represent the
surface $f=0$ and the projection of its apparent contour
(the $x$-direction is pointing to the top of the image and $z$-direction to
the left).
\begin{figure}
 \begin{center}
\img[height=6cm]{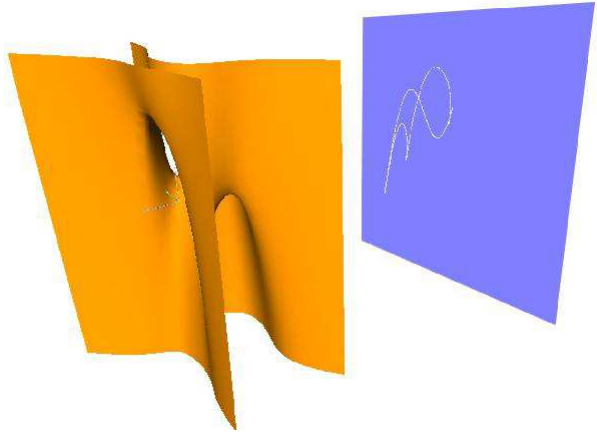} 
\end{center}
\caption{A quartic surface and the projection of a polar curve.}
\end{figure}
\begin{figure}
 \begin{center}
\img[height=6cm]{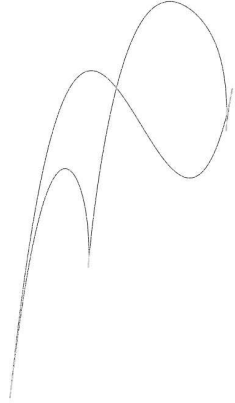}
\end{center}
\caption{The polar curve of the quartic surface projected on a plane.}
\end{figure}
The first factor of degree $10$ has $4$ real roots corresponding to the
smooth points of the surface with a tangent plane orthogonal to the
$x$-direction. The second factor of multiplicity $2$ 
corresponds to points of the polar variety which project in the $(x,y)$-plane
onto the same
point. Geometrically speaking, we have a double folding of the surface
in the $z$-direction above these values. There are two such real points in our
example. The last factor of multiplicity $3$ corresponds to cusp points on
the discriminant curve, which are the projection of 
a ``fronce'' or a pleat of the surface.
There are $4$ such real points. Notice that the branches of the discriminant
curve between two of these cusp points and one of the double folding points
form a very tiny loop, which is difficult to observe at this scale.

These phenomena can be explained from an algebraic point of view, as we will
see hereafter. They have also been analyzed from a singularity theory point
of view. A well-know result in singularity theory, due to H. Whitney (see
\cite{whitney55,mather73,ArVaGo86}) asserts that the singularities of the
projection of a generic surface onto a plane are of $3$ types:
\begin{itemize}
 \item a regular point on the contour curve  corresponding to a fold of the
surface, 
 \item a cusp corresponding to a pleat, 
 \item a double point corresponding to the projection of
two transversal folds.
\end{itemize}
 These are stable singularities, which remain by a small perturbation of the 
surface or of the direction of projection. For a more complete analysis of
the singularities of the apparent contour, see also \cite{platonova86}.

We are now going to analyze the algebraic side of these geometric properties
for generic polynomials of a given degree. Suppose given a positive integer 
$d \geq 3$ and a homogeneous polynomial
\[ \label{Peq} \begin{array}{ll}
     P (X_1, X_2, X_3) \assign \sum_{0 \leq i,  j ; i + j \leq d} &
     U^{}_{i, j} X_1^i X_2^j X_3^{d - i - j} \in \UU[X_1, X_2, X_3],
   \end{array} \]
where $\UU$ is the universal coefficients ring $\mathbbm{Z}[U^{}_{i, j}]$. In this section we will study the discriminant of the discriminant of $P$. Here is the first factorization we can get as a specialization of the iterated resultant we studied in the previous section.

\begin{proposition}\label{discdisc:firstfactor} We have the
  equality in $\UU$:
  \begin{multline*}
    U_{0, 0}^{2\,d^{2}-2 d -1} \Disc_{X_2} (\Disc_{X_3} (P (1,
X_2, X_3)))  =  \Disc_{X_{1}:X_{2}:X_{3}}(P) \times  \\ \Res_{X_{1}:X_{2}:X_{3}} (P, \partial_3
    P, \partial_3^2 P) \times
    \Res_{X_{1}:\cdots:X_{4}} (P , \delta_{3,4} P, \partial_3 P, \delta_{3,4} \partial_3 P).
  \end{multline*}
  The iterated discriminant $\Disc_{X_2} (\Disc_{X_3} (P (1,
  X_2, X_3)))$ is homogeneous with respect to the set of coefficients
  $U^{}_{i, j}$ of degree $4\,(d-1) \,(d^{2}-d-1)$.
\end{proposition}
\begin{proof}
  This is essentially a specialization of the formula given in Proposition
  \ref{DiscRes}, which yields in our case the equality
\begin{multline*} \Disc_{X_2} (\Res_{X_3} (P (1, X_2, X_3), \partial_3 P (1,
    X_2, X_3)))  =   \Disc (P, \partial_3 P ) \times  \\
     \Res(P, \delta_{3,4} P, \partial_3 P, \delta_{3,4} \partial_3 P) .
\end{multline*}
  On the one hand, we know by definition of the discriminant and properties of
  resultants, that
  \begin{eqnarray}
\lefteqn{\Res (P, \partial_3 P, X_1) \Disc (P, \partial_3 P)}\\ & = &
    \Res (P, \partial_3 P, \partial_2 P \partial_3^2 P - \partial_3 P
    \partial_2 \partial_3 P) \nonumber\\
    & = & \Res (P, \partial_3 P, \partial_2 P \partial_3^2 P) \nonumber\\
    & = & \Res (P, \partial_3 P, \partial_2 P) \times 
\Res (P,\partial_3
P,\partial_3^2 P). \label{eq:previous} 
  \end{eqnarray}
Moreover, by Euler identity we have $dP =X_{1}\partial_{1}P+ X_{2}
\partial_{2}P+ X_{3} \partial_{3} P$ from 
we deduce that
\begin{eqnarray*}
{d^{d-1}\Res (P, \partial_3 P, X_1)} 
 & = & \Res(X_{2} \partial_{2}P, \partial_{3} P, X_{1}) \\
 & = &   \Res (X_{2},\partial_{3}P, X_{1})\times \Res(\partial_{2} P,
\partial_{3} P, X_{1})\\
& = &  d\, U_{0,0} \Res(\partial_{2} P, \partial_{3} P, X_{1}).
\end{eqnarray*}
Again by Euler identity, the relation \eqref{eq:discr2} yields
\begin{eqnarray*}
{d^{(d-1)^2}\Res (P, \partial_3 P, \partial_{2}P)}
&=& \Res(X_{1} \partial_{1}P, \partial_{3} P, \partial_{2}P) \\
&=&  \Res(X_{1}, \partial_{3} P, \partial_{2}P) 
\Res(\partial_{1} P, \partial_{3} P, \partial_{2}P) \\
&=& d^{d^{2}-3 d+3} \Res(\partial_{2} P, \partial_{3}P, X_{1}) 
 \Disc(P)
\end{eqnarray*}
and by substitution in \eqref{eq:previous} and  simplification by
$\Res(\partial_{2} P, \partial_{3}P, X_{1})$ and $d^{d-2}$, we get that
$$ 
U_{0,0} \Disc (P, \partial_3 P) = \Disc(P)\Res(P,\partial_3 P,\partial_3^2 P)
$$
so that 
\begin{multline*}
U_{0,0}\Disc_{X_2} (\Res_{X_3} (P (1, X_2, X_3), \partial_3 P (1,X_2,
X_3))) =  
   \Disc(P) \times \\ 
\Res(P,\partial_3 P,\partial_3^2 P)\times \Res_{X_{1}:\cdots:X_{4}} (P (X_3), \delta_{X_3, X_4} (P), \partial_3 P
    (X_3), \delta_{X_3, X_4} (\partial_3 P)).
\end{multline*}
On the other hand, 
\begin{eqnarray*}
\lefteqn{ \Disc_{X_2} (\Res_{X_3} (P (1, X_2, X_3), \partial_3 P (1,
    X_2, X_3)))} \\
& = & \Disc_{X_2} (U_{0, 0} \Disc_{X_3} (P (1,
    X_2, X_3)))\\
    & = & U_{0, 0}^{2\, (d (d - 1) - 1)} \Disc_{X_2}
(\Disc_{X_3}(P (1, X_2, X_3)))\\
    & = & U_{0, 0}^{2 d^{2} -2 d - 2} \Disc_{X_2} (\Disc_{X_3}(P (1, X_2, X_3)))
  \end{eqnarray*}
since $\Disc_{X_3} (P (1, X_2, X_3))$ is of degree $d (d-1)$ in $X_{2}$
and the discriminant of a polynomial of degree $D$ is homogeneous of degree
$2\,(D-1)$ in its coefficients. The claimed formula then follows immediately.
  
Concerning the degree, observe that $\Disc_{X_3} (P(1,X_{2},X_{3}))$ has
coefficients of degree $2\, (d-1)$ in the coefficients of $P$. The
discriminant of a polynomial of degree $D$ being homogeneous of degree 
$2 \, (D-1)$ in its coefficients,  we obtain that $\Disc_{X_{2}}(\Disc_{X_3}
(P(1,X_{2},X_{3})))$
is of degree $$2 (d(d-1)-1)\times 2\,(d-1)= 4(d-1)(d^{2}-d-1)$$ since the
degree of $\Disc_{X_3} (P(1,X_{2},X_{3})))$ in $X_{2}$ is $D=d (d-1)$.
\end{proof}

In the factorization of $\Disc_{X_2}(\Disc_{X_3}(P))$ given in this
proposition, we only know that the factor $\Disc(P)$ is known to be irreducible in
$\UU$. The remaining of this section is devoted to the study of the full
factorization of the two other factors. 
We begin with the study of the factor appearing in Proposition \ref{discdisc:firstfactor} 
which corresponds to the resultant of $P$, its first and second derivatives with
respect to~$X_3$.

\begin{lemma}\label{lm:flex1} Let $\SS$ be a commutative ring and suppose given 
 a linear form $L= a\, X_{1} +b \, X_{2} +c \, X_{3} \in \SS[X_1,X_2,X_3]$
and a homogeneous polynomial $Q \in \SS[X_1,X_2,X_3]$ of degree $d-1\geq 2$. Then 
$$\Res (LQ, \partial_3 (LQ), \partial_3^2 (LQ))= 2^{2(d-1)} c^{3\,d-4}
\Res(L,Q,\partial_{3}(Q))^{3}\times  \Res(Q,\partial_{3}(Q),\partial_{3}^{2}(Q)). 
$$
\end{lemma}
\begin{proof}
It is a straightforward computation using the 
properties of the resultants:
  \begin{eqnarray*}
    \Res (LQ, \partial_3 (LQ), \partial_3^2 (LQ)) &=&  \Res (L
    Q, cQ + L \partial_3(Q), 2 c \partial_3(Q) + L\,
\partial_3^2(Q) )\\
    &=&  \Res (L,c\, Q, 2 c\, \partial_3(Q)) 
\times \\ & & 
     \Res (Q, L\, \partial_3(Q), 2\, c\, \partial_3(Q) + L\, \partial_3^2(Q) )\\
    &= & 2^{d-1} c^{2d-3} \Res (L, Q, \partial_3 (Q)) 
    \Res (Q, L, 2\, c\, \partial_3(Q)) \times \\ 
& & \Res (Q, \partial_3(Q), L\, \partial_3^2(Q))\\
    &=& 2^{2(d-1)} c^{2d-3 + d-1}  \Res (L, Q, \partial_3 Q)^2 \times \\ 
& & \Res (Q,
    \partial_3(Q), L) \Res (Q, \partial_3(Q), \partial_3^2(Q) )\\
    &= &  2^{2(d-1)} c^{3d-4} \Res (L, Q, \partial_3(Q) )^3 \times \\ 
& & \Res (Q,
    \partial_3(Q), \partial_3^2(Q)).
  \end{eqnarray*}
\end{proof}
 
\begin{proposition} \label{prop:discdisc3}
 The  following equality holds in $\UU$:
  \[ \Res_{X_{1}:X_{2}:X_{3}}(P, \partial_3 P,\partial_3^2 P ) = 2^{d (d -
1)} U_{0, 0}^2 \, \Fm(P) \]
  where $\Fm(P)$ is an irreducible homogeneous polynomial in $\UU$ of degree 
$3 d (d - 2)$. 
\end{proposition}
 
\begin{proof}
  We rewrite $P$ as
  \[ P = U_{0, 0} X_3^d + c_1 X_3^{d - 1} + \cdots + c_{d - 2} X_3^2 + c_{d -
     1} X_3 + c_d \]
  where the $c_i$'s are homogeneous polynomials in $\UU[X_1, X_2]$ of degree
  $i$ respectively; \ we have
  \begin{eqnarray*}
    \partial_3 P & = & \tmop{dA}_{0, 0} X_3^{d - 1} + (d - 1) c_1 X_3^{d - 2}
    + \cdots + 2 c_{d - 2} X_3^{} + c_{d - 1,}\\
    \partial_3^2 P & = & d (d - 1) U_{0, 0} X_3^{d - 2} + (d - 1) (d - 2) c_1
    X_3^{d - 3} + \cdots + 2 c_{d - 2} .
  \end{eqnarray*}
  Consider the incidence variety
  \[ \mathcal{W} \assign \{(x_1 : x_2 : x_3) \times (a_{i, j}) \in
     \mathbbm{P^{\mathrm{2}}} \times \mathbbm{A}^{\mathrm{d (d + 1) / 2}}
     \tmop{such} \tmop{that} P = \partial_3 P = \partial_{3^{}}^2 P = 0\} \]
  whose canonical projection onto the second factor, i.e.~by $\pi_2 : \mathcal{W}
  \rightarrow \mathbbm{A}^{\mathrm{d (d + 1) / 2}}$, is the variety of pure codimension
  one defined by the equation $\Res (P, \partial_3 P, \partial_3^2 P) =
  0$. Considering the canonical projection onto its first factor $\pi_1 :
  \mathcal{W} \rightarrow \mathbbm{P}^2$, which is surjective, we observe that
  $\pi_1^{- 1} (0 : 0 : 1)$ is the hyperplane $\{U_{0, 0} = 0\}$ in
  $\mathbbm{A^{\mathrm{d (d + 1) / 2}}}$ and that $\pi_1^{- 1} (x_1 : x_2 :
  x_3)$ is a linear space of codimension 3 in $\mathbbm{A^{\mathrm{d (d + 1) /
  2}}}$ if $(x_1 : x_2 : x_3) \neq (0 : 0 : 1) \in \mathbbm{P}^2$ (just
  observe that in this case the three conditions $c_d=0, c_{d - 1}=0, c_{d - 2}=0$ are non trivial
   and that $\partial_3 P$ does not depend on $c_d$ and that
  $\partial_3^2 P$ does not depend on $c_d$ and $c_{d - 1}$). It follows that
  $\mathcal{W}=\mathcal{W}_0 \cup \mathcal{W}_3$ where $\mathcal{W}_0$ is the
  irreducible variety defined by $\{U_{0, 0} = 0\}$ and $\mathcal{W}_3$, which is the
  closure of the fiber bundle $\pi_1^{- 1} (\mathbbm{P}^2 \setminus (0 : 0 :
  1)) \rightarrow \mathbbm{P}^2 \setminus (0 : 0 : 1)$ in
  $\mathbbm{P^{\mathrm{2}}} \times \mathbbm{A^{\mathrm{d (d + 1) / 2}}}$, is
  an irreducible variety of codimension 3. We deduce that $\pi_2
  (\mathcal{W})$ is the union of two irreducible varieties: $\pi_2
  (\mathcal{W}_0)$ whose defining equation is $\{U_{0, 0} = 0\}$ and $\pi_2
  (\mathcal{W}_3)$. Therefore, we obtain that, in $\UU$,
  \[ \Res (P, \partial_3 P, \partial_3^2 P) = c (d) U_{0, 0}^{a (d)}
     R_d^{r (d)} \]
  where $c (d), a (d), r (d)$ are all positive integers (in particular, they
  are nonzero) which may only depend on $d$, the degree of $P$, and $R_d$ is
  an irreducible polynomial in $\UU$. In order to determine $c (d), a (d), r
  (d)$ we will use a particular specialization and some properties of the
  resultants.
  
  First, it is easy to check by hand (or with a computer), that the case $d =
  3$ gives, in $\UU$,
  \[ \Res (P, \partial_3 P, \partial_3^2 P) = 2^6 U_{0, 0}^2 \Fm(P) \]
  where $\Fm(P)$ is an irreducible polynomial. Moreover, if $\phi$ denotes the
  homogeneous specialization which sends all the coefficients of $c_d$ to 0,
 we have $\phi(P)=X_{3}\, Q$, 
 where $Q$ is a generic homogeneous polynomial of
  degree $d - 1$. By Lemma \ref{lm:flex1}, we get
$$
\phi (\Res (P, \partial_3 P, \partial_3^2 P))  =
2^{2 (d - 1)} \Res (Q, \partial_3 Q, X_3)^3 \Res (Q,\partial_3 Q, \partial_3^2 Q) .
$$
  So, if we proceed by induction on the integer $d$, we obtain
  \begin{eqnarray*}
    \phi (\Res (P, \partial_3 P, \partial_3^2 P)) & = & 2^{2 (d - 1)}
    \Res (Q, \partial_3 Q, X_3)^3 \times 2^{(d - 1) (d - 2)} 
    U_{0,0}^{a (d - 1)} \Fm(Q)\\
    & = & 2^{d (d - 1)} U_{0, 0}^2 \Res (Q, \partial_3 Q, X_3)^3 \Fm(Q)
  \end{eqnarray*}
  where we notice that $\Res (Q, \partial_3 Q, X_3) = \Res (c_{d - 1}, c_{d -
2})$ is an irreducible polynomial in $\UU$ (it is the resultant of two
generic homogeneous polynomials) which does not depend on $U_{0, 0}$. 
  Since we must also have
  \[ \phi (\Res (P, \partial_3 P, \partial_3^2 P)) = c (d) U_{0, 0}^{a
     (d)} \phi (\Fm(P))^{r (d)} \]
  we deduce by comparison that, for all $d \geq 3$,
  \begin{itemizedot}
    \item $c (d)$ divides $2^{d (d - 1)}$,
    
    \item $r (d) = 1$ (since the specialization of $R_d$ by $\phi$ produces an
    irreducible and reduced factor),
    
    \item $a (d) \leq 2$. 
  \end{itemizedot}
  It is easy to see that 2 divides $\partial_3^2 P$ and hence that $2^{d
  (d - 1)}$ divides $\Res (P, \partial_3 P, \partial_3^2 P)$. This
  implies that $2^{d (d - 1)}$ divides $c (d)$ and hence, since we
  already noted that $c (d)$ divides $2^{d (d - 1)}$,  $c (d) = 2^{d (d -
  1)}$. To conclude the proof, it remains to show that $a (d) \geq 2$ for all
  $d \geq 3$, that is to say that $U_{0, 0}^2$ divides $\Res (P,
  \partial_3 P, \partial_3^2 P)$ in $\UU$. Moreover,  it is sufficient to show that
  $U_{0, 0}^2$ divides $$\mathfrak{R} \assign \Res (d (d - 1) P, (d - 1)
  \partial_3 P, \partial_3^2 P)$$ since
  \[ \Res (d (d - 1) P, (d - 1) \partial_3 P, \partial_3^2 P) = d (d -
     1)^{(d - 1) (d - 2)} \times (d - 1)^{d (d - 2)} \times \Res (P,
     \partial_3 P, \partial_3^2 P) . \]
  Using Euler identity several times, we get
  \begin{eqnarray*}
    \mathfrak{R} & = & \Res ((d - 1) X_1 \partial_1 P + (d - 1) X_2
    \partial_2 P, (d - 1) \partial_3 P, \partial_3^2 P)\\
    & = & \Res (X_1^2 \partial_1^2 P + 2 X_1 X_2 \partial_1 \partial_2
    P + X_2^2 \partial_2^2 P, X_1 \partial_1 \partial_3 P + X_2 \partial_2
    \partial_3 P, \partial_3^2 P) .
  \end{eqnarray*}
  It is clear that $\partial_3^2 P \in (X_1, X_2, U_{0, 0} X_3)$ and that  $$X_1
  \partial_1 \partial_3 P + X_2 \partial_2 \partial_3 P \in (X_1, X_2) \subset
  (X_1, X_2, U_{0, 0} X_3).$$ Moreover, $X_1^2 \partial_1^2 P + 2 X_1 X_2
  \partial_1 \partial_2 P + X_2^2 \partial_2^2 P \in (X_1, X_2)^2 \subset
  (X_1, X_2, U_{0, 0} X_3)^2$. Therefore, the divisibility property of resultants implies
  that
  \[ \Res (X_1, X_2, U_{0, 0} X_3)^{2 \times 1 \times 1} \tmop{divides}
     \mathfrak{R}, \]
  and since $\Res (X_1, X_2, U_{0, 0} X_3) = U_{0, 0}$, we are done.

Finally, the degree of $\Fm(P)$ in the coefficients of $P$ is given by the formula
$$ 
d\, (d-1) + d\,(d-2) + (d-1)\,(d-2)-2=3\, d\, (d-2).$$
\end{proof}

We now turn to the study of the third factor appearing in the factorization of $\Disc_{X_2}(\Disc_{X_3}(P))$ in Proposition \ref{discdisc:firstfactor}.

\begin{proposition}\label{prop:dd1}
Assuming that $d\geq 4$, we have the following equality  between homogeneous polynomials in $\UU$ of degree $2 (d - 1) (2 d^2 - 4
  d + 1)$:
 \begin{multline*}
2^{d (d - 1)}\Res_{X_{1}:\cdots:X_{4}}(P, \delta_{3,4}
    P, \partial_3 P, (\delta_{3,4}
    \partial_3) P) = \\
	  \Res_{X_{1}:X_{2}:X_{3}} (P, \partial_3 P,
     \partial_3^2 P)^{2}\times
 \Res_{X_{1}:\cdots:X_{4}} (P , \partial_3 P,\delta_{3,4}^{2} P,
(\delta_{3,4}^{2}\partial_{3} -2 \delta_{3,4}^{3}) P).
  \end{multline*}
\end{proposition}
\begin{proof}
Using known properties of resultants and equalities \eqref{eq1} and \eqref{eq2}, we get the equalities
  \begin{eqnarray*}
    \Res (P, \delta_{3,4} P, \partial_3 P, \delta_{3,4}\partial_3 P) & = &
    \Res (P, \partial_3 P + (X_4 - X_3) \delta_{3,4}^{2}P, \partial_3 P, \delta_{3,4}\partial_3 P)\\
    & = & \Res (P, (X_4 - X_3) \delta_{3,4}^{2}P, \partial_3 P, \delta_{3,4}\partial_3 P)\\
    & = & \Res (P, (X_4 - X_3), \partial_3 P, \delta_{3,4}\partial_3 P) \times\\
& &  
    \Res (P, \delta_{3,4}^{2}P, \partial_3 P, \partial_3^2 P + (X_4 -
X_3) \delta_{3,4}^{2}\partial_{3}P)\\
    & = & \Res (P, \partial_3 P, \partial^{2}_3 P)\times\\
& &  
    \Res (P, \delta_{3,4}^{2}P, \partial_3 P, \partial_3^2 P + (X_4 - X_3) \delta_{3,4}^{2}\partial_{3}P).
  \end{eqnarray*}
Let us denote $$\mathcal{S}(P):=\Res (P, \delta_{3,4}^{2}P, \partial_3 P, \partial_3^2 P +
    (X_4 - X_3) \delta_{3,4}^{2} \partial_{3}P).$$
Since $\partial_{3}^{2} P = 2 \delta_{3,4}^{2} P- 2 (X_{4}-X_{3})\delta_{3,4}^{3}P$ in $\UU$, we deduce that, in $\UU$,
  \begin{eqnarray*}
2^{d(d-1)(d-2)}\mathcal{S}(P) & =  & \Res (P, 2\delta_{3,4}^{2}P, \partial_3 P, \partial_3^2 P +
    (X_4 - X_3) \delta_{3,4}^{2} \partial_{3}P) \\
     & = &  \Res (P, 2\delta_{3,4}^{2}P, \partial_3 P, (X_4 - X_3)
(\delta_{3,4}^{2}\partial_{3} - 2\delta_{3,4}^{3})P)\\
     & = &  \Res (P, 2\delta_{3,4}^{2}P, \partial_3 P, (X_4 - X_3))
\times \\ & & \Res (P, 2\delta_{3,4}^{2}P, \partial_3 P,
(\delta_{3,4}^{2}\partial_{3} - 2\delta_{3,4}^{3})P)\\
     & = &  \Res (P, \partial_3^{2}P, \partial_3 P, (X_4 - X_3))
\times \\ & & \Res (P, 2\delta_{3,4}^{2}P, \partial_3 P,
(\delta_{3,4}^{2}\partial_{3} - 2\delta_{3,4}^{3})P)\\
     & = &  2^{d(d-1)(d-3)}\Res (P, \partial_{3}^{2}P, \partial_3 P)
\times   \\ & & \Res (P, \delta_{3,4}^{2}P, \partial_3 P,
(\delta_{3,4}^{2}\partial_{3} - 2\delta_{3,4}^{3})P).
  \end{eqnarray*}
  Therefore, we have
  \begin{multline*} 2^{d (d - 1)} \Res (P, \delta_{3,4} P, \partial_3 P, (\delta_{3,4}
     \partial_3) P) =  \Res (P, \partial_3 P, \partial_3^2 P)^2
\times \\ \Res ((P, \partial_{3}P, \delta_{3,4}^{2}P, 
(\delta_{3,4}^{2}\partial_{3} - 2\delta_{3,4}^{3})P). \end{multline*}
  The claimed formula for the degree of $\Res (P, \delta_{3,4} P, \partial_3
  P, (\delta_{3,4} \partial_3) P)$ follows immediately by specialization of the
  formula given in Lemma \ref{squareres}.
\end{proof}

As a consequence, to get the full factorization of $\Disc_{X_2}(\Disc_{X_3}(P))$ it only remains to study the factorization of the term
$$ 
\Rc(P) 
:= \Res _{X_{1}:\cdots:X_{4}}(P, \partial_3 P,\delta_{3,4}^{2}P,(\delta_{3,4}^{2}\partial_{3} -2 \delta_{3,4}^{3})P)
$$
whose degree in the coefficients of $P$ is 
$$ 
2\,(d-1)\,(2d^{2}-4d+1) - 2 (d\, (d-1)+ d\, (d-2) + (d-1)\,(d-2)) = 2\, (2d-3)\,(d^{2}-3d+1). 
$$
This is the aim of the next proposition. We begin with two technical
lemmas. 
\begin{lemma}\label{lm:dd2} Let $\SS$ be a commutative ring and suppose given 
 a linear form $L= a\, X_{1} +b \, X_{2} +c \, X_{3} \in \SS[X_1,X_2,X_3]$ and a homogeneous polynomial $Q
\in \SS[X_1,X_2,X_3]$ of degree $d-1\geq 2$. Then 
\begin{multline*}
\Res_{X_{1}:\cdots:X_{4}} (LQ, \partial_3 (LQ) ,
     \delta_{3,4}(LQ),
     \delta_{3,4}\partial_{3} (LQ)) = 2^{2(d-1)}c^{2(d-1)(3d-5)} \times \\ \Res_{X_{1}:X_{2}:X_{3}}(L,Q,\partial_{3}
Q)^{6} \times   \Res_{X_{1}:\cdots:X_{4}} (L(X_{4}),Q,\partial_{3} Q,
\delta_{3,4}^{2}Q)^{4}  \times  \\ \Res _{X_{1}:\cdots:X_{4}}(Q,\delta_{3,4} Q, \partial_{3} Q, \delta_{3,4} \partial_{3} Q).
\end{multline*}
\end{lemma}
\begin{proof} Set $P:=LQ$. 
Applying \eqref{eq:dr1} with $P_{2} =\partial_{3}(P)$ and
$$\partial_{3} P(X_{4}) = \partial_{3} P(X_{3})+ (X_{4}-X_{3})\, \delta_{3,4}\partial_{3} P,$$ we
obtain the decomposition 
\begin{eqnarray}\label{eq:proddd}
\lefteqn{\Res (P , \partial_{3}(P), \delta_{3,4} (P), \delta_{3,4}\partial_{3}(P))}\\
&=&  c^{(d-1)(d-2)}\, 
\times \Res (L(X_{4}), Q(X_3),\partial_{3}(P), \delta_{3,4} \partial_{3} (P))^{2}
\times \nonumber \\ & & \Res (Q, \delta_{3,4}(Q),\partial_{3}(P), \delta_{3,4} \partial_{3}(P))
\nonumber\\
& = &   c^{(d-1)(d -2)}\, R_{1}^{2}\, R_{2}.  \nonumber
\end{eqnarray}
Using the relations
\begin{eqnarray*}
\partial_{3} (P) &= & \partial_{3}(L)\, Q + L\, \partial_{3} (Q), \\
\delta_{3,4}\partial_{3} (P) &= & \partial_{3}(L)\, \delta_{3,4}(Q) + \delta_{3,4}(L)\, \partial_{3} (Q) +
L(X_{4}) \delta_{3,4} \partial_{3}(Q),
\end{eqnarray*}
we can decompose further the previous expressions:
\begin{eqnarray*}
R_{1} := 
\lefteqn{\Res (L(X_{4}), Q,\partial_{3}(P), \delta_{3,4} \partial_{3}(P) )}\\
& = & \Res (L(X_{4}), Q,L\, \partial_{3}(Q), \partial_{3}(L)\,
\delta_{3,4}(Q) +\delta_{3,4}(L)\, \partial_{3}(Q))\\
& = & \Res (L(X_{4}), Q,L, c\,\delta_{3,4}(Q) + c\, \partial_{3}(Q))
\times \Res (L(X_{4}), Q,\partial_{3}(Q), c\,\delta_{3,4}(Q))\\
& = & 
\Res (L(X_{4}), Q, c\, (X_{3}-X_{4}), c \, (\delta_{3,4}(Q) +\partial_{3}(Q)))
\times \\
& & c^{(d-1)(d-2)}\Res (L(X_{4}), Q, \partial_{3}(Q), \delta_{3,4}(Q))\\
& = & c^{(d-1)+ (d-1)(d-2)}
\Res (L, Q, 2  \partial_{3}(Q))
\times \\ & & c^{(d-1)(d-2)}\Res (L(X_{4}), Q, \partial_{3}(Q), \delta_{3,4}(Q))\\
& = & 2^{d-1}c^{(d-1)(2 d -3)}
\Res (L, Q, \partial_{3}(Q)) 
\times \Res (L(X_{4}), Q, \partial_{3}(Q), \delta_{3,4}(Q)).
\end{eqnarray*}
Now, using the relation $\delta_{3,4}(Q)=\partial_{3}(Q) + (X_{4}-X_{3})\delta_{3,4}^{2}(Q)$, we have
\begin{eqnarray}\label{eq:proddd2}
\lefteqn{\Res (Q,\partial_{3}(Q),\delta_{3,4}(Q),L(X_{4}))} \\ 
&=& \Res(Q,\partial_{3}(Q),(X_{4}-X_{3})\,\delta_{3,4}^{2}(Q),L(X_{4})) \nonumber \\
&=& \Res(L, Q,\partial_{3}(Q))
\times \Res(Q,\partial_{3}(Q), \delta_{3,4}^{2}(Q),L(X_{4})). \nonumber
\end{eqnarray}
We deduce that
$$ 
R_{1} =   2^{d-1} c^{(d-1)(2 d -3)} \Res (L, Q, \partial_{3}(Q))^{2}
\times \Res (Q, \partial_{3}(Q), \delta_{3,4}^{2}(Q), L(X_{4})).
$$
Similarly, we have
\begin{eqnarray*}
R_{2} =\lefteqn{\Res (Q, \delta_{3,4}(Q), \partial_{3}(P), \delta_{3,4} \partial_{3}(P))}\\ 
&=& \Res (Q, \delta_{3,4}(Q), L\,\partial_{3}(Q) ,
\delta_{3,4}(L) \partial_{3}(Q) + L(X_{4})\, \delta_{3,4}\partial_{3} (Q))\\
&=&
\Res (Q, \delta_{3,4}(Q), L, \delta_{3,4}(L)
\partial_{3}(Q) + L(X_{4}) \,\delta_{3,4}\partial_{3} (Q))
\times \\ & &
\Res (Q, \delta_{3,4}(Q), \partial_{3}(Q), L(X_{4})\, \delta_{3,4}\partial_{3} (Q))\\
&=&
\Res (Q, \delta_{3,4}(Q), L, c\, \partial_{3}(Q)(X_{4}))
\times 
\Res (Q, \delta_{3,4}(Q), \partial_{3}(Q), L(X_{4}))\\
&&\times 
\Res (Q, \delta_{3,4}(Q), \partial_{3}(Q), \delta_{3,4}\partial_{3} (Q))\\
&=& 
c^{(d-1)(d-2)}
\Res (Q, \partial_{3}(Q),  \delta_{3,4}(Q), L(X_{4}))^{2}
\times \\ & &
\Res (Q, \delta_{3,4}(Q), \partial_{3}(Q), \delta_{3,4}\partial_{3}(Q) ),
\end{eqnarray*}
since 
$$ 
c \partial_{3}(Q)  + L(X_{4}) \delta_{3,4} \partial_{3} Q - L (X_{3})\,\delta_{3,4} \partial_{3} (Q) 
= c \partial_{3}(Q)(X_{4}),
$$
and 
$$
\Res (Q, \delta_{3,4}(Q), L, \partial_{3}(Q)(X_{4}))
=
\Res (Q, \partial_{3}(Q),  \delta_{3,4}(Q), L(X_{4})).
$$
using the relation \eqref{eq:proddd2}, we deduce that
\begin{multline*}
R_{2} =   c^{(d-1)\, (d-2)} \Res(L, Q,\partial_{3} Q)^{2} \times \Res(L(X_{4}), Q, \partial_{3} Q,
\delta_{3,4}^{2}Q)^{2}
\times \\
\Res (Q, \delta_{3,4}(Q), \partial_{3}(Q), \delta_{3,4}\partial_{3}(Q) ).
\end{multline*}
By \eqref{eq:proddd}, squaring $R_{1}$ and taking the product with
$R_{2}$ and $c^{(d-1)(d-2)}$, we obtain the expected decomposition.
\end{proof}

\begin{lemma}\label{lm:dd3} Let $\SS$ be a commutative ring and suppose given 
 a linear form $L= a\, X_{1} +b \, X_{2} +c \, X_{3} \in \SS[X_1,X_2,X_3]$ and a homogeneous polynomial $Q
\in \SS[X_1,X_2,X_3]$ of degree $d-1\geq 2$. Then 
$$ 
\Rc(LQ)=  c^{6d^2-22d+18}\, \Res (L(X_{4}),Q , \partial_{3}(Q), \delta_{3,4}^{2}(Q))^{4}\,
\Rc(Q).
$$
\end{lemma}
\begin{proof}
Set $P:=LQ$. By  Proposition \ref{prop:dd1} and Lemma \ref{lm:flex1}, we have
\begin{eqnarray*}
\lefteqn{2^{d (d-1)} \Res (P , \delta_{3,4} (P) , \partial_{3}(P) , \delta_{3,4}\partial_{3}(P) )} \\  
 &=&    \Res (P, \partial_{3}(P), \partial_{3}^{2}(P))^{2} \, \Rc(P) \\
 &=&   \left(2^{2(d-1)} c^{3d-4}
\Res(L,Q,\partial_{3}(Q))^{3}\, \Res(Q,\partial_{3}(Q),\partial_{3}^{2}(Q))
\right)^{2}\Rc (P)\\
 &=&   2^{4(d-1)} c^{6d-8}
\Res(L,Q,\partial_{3}(Q))^{6}\, \Res(Q,\partial_{3}(Q),\partial_{3}^{2}(Q))^{2}\, \Rc (P).
\end{eqnarray*}
Moreover, by Lemma \ref{lm:dd2} and Proposition \ref{prop:dd1} we also have
\begin{eqnarray*}
\lefteqn{\Res (P , \delta_{3,4} (P) , \partial_{3}(P) , \delta_{3,4}\partial_{3}(P) )} \\
 & = &  
2^{2(d-1)}c^{2(d-1)(3d-5)} \Res
(L,Q,\partial_{3} Q)^{6}   \, \Res (L(X_{4}),Q,\partial_{3} (Q), \delta_{3,4}^{2}(Q))^{4} 
\times \\ & & \Res (Q,\delta_{3,4} Q, \partial_{3} Q, \delta_{3,4}\partial_{3} Q)\\
&=& 2^{-d(d-1)}c^{2(d-1)(3d-5)} \Res(L,Q,\partial_{3} Q)^{6}  \,
 \Res (L(X_{4}),Q,\partial_{3} (Q), \delta_{3,4}^{2}(Q))^{4} 
\times \\ & & 
\Res(Q,\partial_{3} Q, \partial_{3}^{2}Q)^{2} \Rc(Q)
\end{eqnarray*}
and the claimed formula follows by comparison.
\end{proof}

\begin{proposition} \label{prop:discdisc5}
Assuming that $d\geq 4$, we have  
$$ 
\mathcal{R}(P)= U_{0,0}^{2 d (d-1)-6} \Um(P)^{2},
$$
where $\Um(P)$ is irreducible in $\UU$ of degree $2\,d\,(d-2)\,(d-3)$. 
\end{proposition}
\begin{proof} We first prove that $U_{0,0}^{2d(d-1)-6}$  divides $\Rc(P)$ in $\UU$. 
Gathering the results of the propositions \ref{discdisc:firstfactor}, \ref{prop:discdisc3},
\ref{prop:dd1} and \ref{prop:discdisc5} we obtain the following equality in $\UU$:
\begin{align}\label{almeq}
U_{0,0}^{2d(d-1)-6}\Disc_{X_2}(\Disc_{X_3}(P(1,X_2,X_3)))=U_{0,0}\Disc_{X_{1}:X_{2}:X_{3}}(P)\Fm(P)^3\Rc(P).
\end{align}
In order to prove that $U_{0,0}^{2d(d-1)-6}$ divides $\Rc(P)$, it is thus
sufficient to prove that $U_{0,0}$ divides
$\Disc_{X_2}(\Disc_{X_3}(P(1,X_2,X_3)))$.  Rewrite the polynomial $P$ as
$$P:=U_{0,0}X_3^d+a_{d-1}X_3^{d-1}+\cdots+a_1X_3+a_0$$
where the $a_i$'s are homogeneous polynomials in $X_1,X_2$. 
If one specializes $U_{0,0}$ to 0 then $P$ specializes to a polynomial of degree $d-1$ in $X_3$ and hence, by a well-known property of discriminants we have
$$\Disc_{X_3}(P(1,X_2,X_3))=(-1)^da_{d-1}(1,X_2)^2\Disc'_{X_3}(P(1,X_2,X_3))$$ 
where $\Disc'_{X_3}(P(1,X_2,X_3))$ denotes the discriminant of $P$ as a
polynomial of degree $d-1$ (and not $d$). But then, the discriminant with
respect to $X_2$ of the above quantity equals 0 since it contains a square
factor. This implies that $U_{0,0}$ divides
$\Disc_{X_2}(\Disc_{X_3}(P(1,X_2,X_3)))$.

\medskip

Observe now that by Proposition \ref{squareres} and Proposition \ref{prop:dd1}, 
$\Rc(P)$ is a square and hence can be decomposed
as $$\Rc(P)= c(d)^2 U_{0,0}^{2d(d-1)-6}\,\Rm_{1}^{2\,r_{1}(d)}(P)\, \Rm_{2}^{2\,r_{2}(d)}(P)\cdots
\Rm_{s}(P)^{2\,r_{s}(d)},$$ where $\Rm_{i}(P)$ are irreducible polynomials and $c(d)\in \ZZ$.
As we did several times, we will prove the claimed factorization by induction
on the degree $d$ of $P$. For $d=4$, we find by explicit computation that
$$ \Rc(P):=2^8U_{0,0}^{18}\Um(P)^2$$
where $\Um(P)$ is an irreducible polynomial in $\UU$ of the expected degree.
Assume that for a generic polynomial $Q$ such that $\deg(Q)<d$, we have 
$$\Rc(Q)= U_{0,0}(Q)^{2 (d-1)(d-2) -6} \Um(Q)^{2}$$ 
where
$\Um(Q)$ is an irreducible polynomial of bi-degree $(0, 2(d-1)(d-3)(d-4))$
in $(L,Q)$.

Consider the specialization $\phi(P)=L\, Q$ where $L$ and $Q$ are generic
polynomials of degree $1$ and $(d-1)$ respectively. We denote by $c$, resp.~$U_{0,0}'$, 
the coefficient of $X_3$, resp.~$X_3^{d-1}$, of $L$, resp.~$Q$.
Each factor $\phi(\Rm_{i})$  must decompose into a product of
irreducible factors such that the degree in the coefficients of $L$ is equal to 
the degree in the coefficients of $Q$. By Lemma \ref{lm:dd3}, we have 
$$ 
\Rc(LQ)= c^{6d^2-22d+18}\, \Res (L(X_{4}),Q ,\partial_{3}(Q), \delta_{3,4}^{2}(Q))^{4}
\Rc(Q) 
$$
and by Proposition \ref{calc-lemme}, we have 
$$
\Res (L(X_{4}),Q, \partial_{3}(Q),\delta_{3,4}^{2}(Q)) = U_{0,0}'^{d-1} \Tm(Q,L)
$$
 where $\Tm(Q,L)$
is irreducible of degree $((d-1)(d-2)(d-3), (3 d-4)(d-3))$ in $(L,Q)$.
Using now the induction hypothesis, we deduce that
\begin{eqnarray*}
\Rc(LQ) &=&c^{6d^2-22d+18} U_{0,0}'^{4(d-1)+2(d-1)(d-2)-6} \Tm(Q,L)^{4} \Um(Q)^{2}\\
&=& c^{6d^2-22d+18} U_{0,0}'^{2 d(d-1)-6} \Tm(Q,L)^{4} \Um(Q)^{2} \\
&=& \left(cU_{0,0}'\right)^{2d(d-1)-6}c^{4(d-2)(d-3)}\Tm(Q,L)^{4} \Um(Q)^{2}.
\end{eqnarray*}
An explicit analysis, as the ones we did several times before in this paper, shows that the product
$c^{2(d-2)(d-3)}\Tm(Q,L)^{2} \Um(Q)$ must comes from the same factor
$\Rm_{i}(P)$. Indeed, counting 1 for the degree of a coefficient of $L$ and
$-1$ for the degree of a coefficient of $Q$, the degree of a term $c^{l}
\Tm(Q,L)^{k} \Um(Q)$ with $l\le 2 (d-2)(d-3)$ is at  most
\begin{multline*}
2(d-2)(d-3) + k ( (d-1)(d-2)(d-3)-(3d-4)(d-3))  - 2(d-1)(d-3)(d-4) \\ 
 =  2 (d-3) (d^{2} - 6 d+6) (k-2).
\end{multline*}
which is $< 0$ for $k=0,1$ with $d\ge 5$ and $0$ for $k=2$. 
This proves that for $d\ge 5$,
any sub-product $c^{2(d-2)(d-3)}\Tm(Q,L)^{2} \Um(Q)$ 
has not the same degree in the coefficients of $L$ and $Q$, except when $k=2$ for the product
of all terms. We deduce that $\Um(P)$ is irreducible, which concludes the
proof of the proposition. 
\end{proof}
 
Gathering all our results, we get the following theorem:
\begin{theorem} Assuming that $d\geq 4$, 
we have the equality in $\UU$ 
$$    \Disc_{X_2} (\Disc_{X_3} (P (1,
    X_2, X_3)))  =   U_{0,0}\, \Disc_{X_{1}:X_{2}:X_{3}}(P)\, \Fm(P)^3\, \Um(P)^2.$$
  The iterated discriminant $\Disc_{X_2} (\Disc_{X_3} (P (1,
  X_2, X_3)))$ is homogeneous with respect to the set of coefficients
  $U_{i, j}$ of degree $4\,(d-1) \,(d^{2}-d-1)$.
\end{theorem}
\begin{proof} It is a direct consequence of \eqref{almeq} and
Proposition \ref{prop:discdisc5}. 
\end{proof}

\begin{corollary}
  \label{discdisc1}Given a polynomial $f (\mathbf{x}, y, z)$ of the form
  \[ f (\mathbf{x}, y, z) = \sum_{| \alpha | + i + j \leqslant d}
     a_{\alpha,i, j} \mathbf{x}^{\alpha} y^i z^j \in \SS[\mathbf{x}]
     [y, z], \]
  where $\mathbf{x}$ denotes a set of variables $(x_1, \ldots, x_n)$ for
  some integer $n \geq 1$ and $\SS$ is any commutative ring, the iterated
  discriminant $\Disc_y (\Disc_z (f)) \in \SS [\mathbf{x}]$ has
  degree at most $d (d - 1) (d^2 - d - 1)$ in $\mathbf{x}$.
  If the polynomial $f$ is sufficiently generic and $\SS$ is an infinite field,
we have
$$ \Disc_y (\Disc_z (f))  =  
a_{0, 0, d}\,  \Disc_{y, z} (f)\, \Fm(f)^{3}\, \Um(f)^{2}.$$
where  
\begin{itemizedot}
  \item $\Disc_{y, z} (f)$ is an irreducible polynomial in
    $\mathbf{x}$ of degree $d (d - 1)^2$,
    
  \item $\Fm(f)$ is irreducible in $\mathbf{x}$ of degree $d (d-1)(d - 2)$
and we have the equality $\Res_{y, z} (f, \partial_z f, \partial_z^2 f)=
2^{d(d-1)}\,a_{0,0,d}^{2}\, \Fm(f).$ 

    \item  $\Um(f)$ is an irreducible polynomial in $\mathbf{x}$ of degree
$\frac{1}{2}\, d (d - 1) (d - 2) (d-3)$,
    such that 
$$\Res_{y, z, z'} (f, \delta_{z, z'} (f), \partial_z f, \delta_{z,
z'}(\partial_z f)) = a_{0,0,d}^{4} \Fm(f)^{2} \Um(f)^{2}$$
and also
$$ \Res_{y,z,z'}(f, \partial_z f,\delta^{2}_{z,z'}(f),(\delta_{z,z'}^{2}\partial_{3} -2 \delta_{z,z'}^{3})f)=a_{0,0,d}^{2d(d-1)-6}\Um(f)^{2}.$$
  \end{itemizedot} 
\end{corollary}

\begin{remark}
In the decomposition formula given in the above corollary, we can replace
$\Disc_{y, z} (f)$ by a resultant up to the constant factor $a_{0,0,d}^{2}$.
\end{remark}

\bigskip

\noindent{\bf Acknowledgments:} This work was partially supported by the french ANR GECKO.

\def\cprime{$'$}
\providecommand{\bysame}{\leavevmode\hbox to3em{\hrulefill}\thinspace}
\providecommand{\MR}{\relax\ifhmode\unskip\space\fi MR }
\providecommand{\MRhref}[2]{%
  \href{http://www.ams.org/mathscinet-getitem?mr=#1}{#2}
}
\providecommand{\href}[2]{#2}

\end{document}